\documentclass[12pt,pdfstartview=FitH]{article}

\usepackage{graphicx} 
\usepackage{amssymb}
\usepackage{amsmath}
\usepackage{jheppub}
\usepackage{empheq}
\usepackage{bbm}
\usepackage{enumitem}
\usepackage{physics}
\usepackage{xcolor}
\usepackage{empheq}
\usepackage{tikz}
\usepackage{picture}
\usepackage[Smaller,makeroom,thicklines]{cancel}
\usepackage{Shortcuts_Kaustubh}

\title{Complex Kerr-AdS Black Holes}
\author{Kaustubh Singhi}
\affiliation{International Centre for Theoretical Sciences (ICTS-TIFR), \\ Tata Institute of Fundamental Research, \\ Shivakote, Hesaraghatta, Bangalore 560089, India.}
\emailAdd{kaustubh.singhi@icts.res.in}

\date{\today}

\abstract{We revisit thermodynamics of five-dimensional AdS spacetime at finite temperature and rotation using the Euclidean path integral. It is generally believed that at low temperatures and finite rotation, the bulk saddle point that governs the thermodynamics describes a rotating gas of thermal radiation.
Consequently, the dual gauge theory at low temperatures is in a confined thermal state.
We demonstrate that this holographic expectation is at odds with the fact that, even at low temperatures, there exist saddles of the bulk path integral with real part of on-shell action smaller than that of the thermal rotating gas.
The usual Kerr-AdS black holes but with complex parameters are examples of such saddles.
Using mini-superspace ideas and steepest descent, we argue that these additional saddles do not actually feature in the low temperature partition function.
This saves the original claim that rotating thermal gas is indeed the correct background for understanding the dual gauge theory at low temperatures.
As a corollary, we also find that the unstable small rotating black hole does not contribute to the partition function at any temperature, even in a suppressed manner.
}

\begin{document}
\maketitle

\section{Introduction}
Black holes can be thought of as statistical systems with a temperature and thermodynamics associated to them \cite{Bardeen:1973gs}. One way to arrive at this conclusion is through the Euclidean path integral introduced by Gibbons and Hawking \cite{Gibbons:1976ue} (see section 6 of \cite{Witten:2024upt} for a recent review). The idea is that certain partition functions are computable by performing a path integral over Euclidean geometries while keeping the induced metric on a codimension one asymptotic boundary fixed. If a black hole solution (saddle) contributes and dominates the path integral then the thermodynamics of the solution can be obtained from the partition function using usual statistical methods. \\

A general path integral can have multiple saddles contributing in which case more can be deduced about the system.
An important example is due to Hawking and Page \cite{Hawking:1982dh}, who analyzed the path integral with boundary conditions induced from a Euclidean AdS-Schwarzschild black hole.
This path integral computes the partition function at a temperature fixed by the black hole parameters.
A peculiar feature of AdS spacetimes was realized in \cite{Hawking:1982dh} --- there is a critical temperature corresponding to a phase transition between thermal AdS and the spherical black hole geometry.
This point is commonly referred to as the Hawking-Page phase transition.
From the perspective of the path integral, the phase transition is simply an exchange of dominance between the thermal and black hole saddles as the temperature is varied.
A similar result can be derived for the case with rotation.
There is a Hawking-Page phase transition between a rotating thermal gas in AdS and the rotating black hole geometry.
This is captured by the partition function computed from a Euclidean path integral with boundary conditions set by a Kerr-AdS black hole. \\

Soon after the original proposal of the AdS/CFT correspondence \cite{Maldacena:1997re,Gubser:1998bc,Witten:1998qj}, it was argued in \cite{Witten:1998zw} that the phase transition in the bulk spacetime should be understood as a confinement-deconfinement phase transition in the boundary theory.
Through semi-classical computations in the thermal AdS and black hole backgrounds, the correspondence has revealed interesting features of the dual gauge theory in the confined and deconfined phases, respectively. \\

Recently, it was pointed out in \cite{Mahajan:2025bzo} for the Schwarzschild case that there is a gap in the entire reasoning above stemming from the fact that at lower temperatures the black holes exist as complex saddles of the path integral.
In fact, at very low temperatures, the real part of the on-shell action of these \textit{complex black holes} is smaller than that of the thermal AdS solution.\footnote{This was pointed out in the five-dimensional case, but it is easy to check that the same is true in all higher dimensions as well. Hence, the situation in five and higher dimensions is somewhat different from the more frequently studied four-dimensional case (see for example \cite{Jonas:2022uqb}; appendix F of \cite{Maldacena:2019cbz}).}
Na\"ively including these saddles lead to an inconsistency with the expectation that the dual gauge theory is confined at very low temperatures \cite{Aharony:2003sx}.
The fact that these saddles are complex does not a priori mean that we can discard them. Indeed, complex geometries have proven useful in quantum cosmology \cite{Hertog:2011ky,Feldbrugge:2017kzv,Narain:2021bff,Jonas:2022uqb,Maldacena:2024uhs,Ivo:2024ill,Turiaci:2025xwi,Shi:2025amq,Ivo:2025yek}, gravitational description of quantum chaos \cite{Saad:2018bqo,Saad:2019pqd,Stanford:2020wkf,Chen:2023hra}, real-time holography \cite{Skenderis:2008dh,Glorioso:2018mmw}, as well as computing supersymmetric indices \cite{Boruch:2023gfn,Hegde:2024bmb,Adhikari:2024zif,Boruch:2025biv}. \\

The first objective of this paper is to show the existence of complex black hole saddles in five-dimensional AdS spacetime that can contribute to the ensemble at finite temperature and rotation.
Analogous to the Schwarzschild case, the continuation of the small and large Kerr-AdS black holes to complex parameters are examples of such complex saddles at low temperatures.
We also find additional saddles at finite rotation which do not have an analogue in the case without rotation.
The remainder of this paper is devoted towards restoring the claim that at lower temperatures and finite rotation, the correct saddle that captures the dual gauge theory is indeed the rotating thermal AdS solution. \\

The plan of the paper is as follows. In section \ref{section: Kerr thermodynamics 1}, we collect classic results for the five-dimensional Kerr-AdS black hole with a single rotation.
The thermodynamic properties of this solution is reviewed in section \ref{section: Thermodynamics and Hawking-Page}.
We identify the complex black holes that can contribute to the partition function at finite temperature and rotation in section \ref{section: complex Kerr-AdS solutions}.
In section \ref{section: minisuperpsace}, we use mini-superspace methods and Picard-Lefshetz theory to determine which saddles actually contribute to the partition function at different values of the thermodynamic variables.
Finally, we conclude in section \ref{section: Discussion} with a brief discussion.


\section{Rotating AdS Black Holes in Five Dimensions}
\label{section: Kerr thermodynamics 1}
We start by reviewing the standard thermodynamical relations for a rotating black hole. The Lorentzian metric of a five-dimensional Kerr-AdS black hole with a single rotation is conveniently written as \cite{Hawking:1998kw,Gibbons:2004ai}
\begin{align}
    \begin{split}
        \label{Kerr metric 1}
        ds^{2} & = -\frac{\Delta_{r}}{\rho^{2}}\l(dt - \frac{a}{\Xi}\sin^{2}\theta d\phi\r)^{2} + \frac{\rho^{2}}{\Delta_{r}}dr^{2} + \frac{\rho^{2}}{\Delta_{\theta}}d\theta^{2} \\
        & \qquad + \frac{\Delta_{\theta}\sin^{2}\theta}{\rho^{2}}\l(adt - \frac{r^{2} + a^{2}}{\Xi}d\phi\r)^{2} + r^{2}\cos^{2}\theta d\psi^{2}.
    \end{split}
\end{align}
Here, the $\theta$ coordinate ranges from $0$ to $\frac{\pi}{2}$ while the $\phi,\psi$ coordinates are $2\pi$ periodic. The functions appearing in \eqref{Kerr metric 1} are defined as
\begin{gather}
    \label{Kerr metric 2}
    \Delta_{r}(r) = (r^{2} + a^{2})(r^{2} + 1) - (r_{+}^{2} + a^{2})(r_{+}^{2} + 1), \\
    \label{Kerr metric 3}
    \rho^{2}(r,\theta) = r^{2} + a^{2}\cos^{2}\theta, \quad \Delta_{\theta}(\theta) = 1 - a^{2}\cos^{2}\theta, \quad \Xi = 1 - a^{2}.
\end{gather}
We have set the AdS length to unity. Going to Euclidean time ($\tau = it$) gives the following metric
\begin{gather}
    \label{Kerr metric 4}
    ds^{2} = \frac{\Delta_{r}}{\rho^{2}}\l(d\tau - i\frac{a}{\Xi}\sin^{2}\theta d\phi\r)^{2} + \frac{\rho^{2}}{\Delta_{r}}dr^{2} + \frac{\Delta_{\theta}\sin^{2}\theta}{\rho^{2}}\l(iad\tau + \frac{r^{2} + a^{2}}{\Xi}d\phi\r)^{2},
\end{gather}
where the $\theta$ and $\psi$ directions have been suppressed. For real $r_{+}$ and $\abs{a} < 1$, the metric \eqref{Kerr metric 4} is called quasi-Euclidean. \\

To identify the inverse temperature ($\beta$) and the angular velocity ($\O$) that enter the thermodynamics, we need to analyze the near horizon metric. Taking $r \to r_{+}$ in \eqref{Kerr metric 4} yields
\begin{align}
    \begin{split}
        \label{Kerr metric 5}
        ds^{2} & = \frac{(r - r_{+})\Delta_{+}'}{\rho_{+}^{2}}\l(d\tau - i\frac{a}{\Xi}\sin^{2}\theta d\phi\r)^{2} \\
        & \qquad + \frac{\rho_{+}^{2}}{(r - r_{+})\Delta_{+}'}dr^{2} + \frac{\Delta_{\theta}\sin^{2}\theta}{\rho_{+}^{2}}\l(iad\tau + \frac{r_{+}^{2} + a^{2}}{\Xi}d\phi\r)^{2},
    \end{split}
\end{align}
where we have defined $\Delta_{+}' \equiv \eval{\frac{d\Delta_{r}}{dr}}_{r = r_{+}}$ and $\rho_{+} \equiv \rho(r_{+})$. We define the following angular coordinate
\begin{gather}
    \label{Kerr metric 6}
    \wh{\phi} = \phi + i\frac{a\Xi}{r_{+}^{2} + a^{2}}\tau
\end{gather}
and write \eqref{Kerr metric 5} as
\begin{align}
    \begin{split}
        \label{Kerr metric 7}
        ds^{2} & = \frac{(r - r_{+})\Delta_{+}'\rho_{+}^{2}}{(r_{+}^{2} + a^{2})^{2}}\l(d\tau - i\frac{a(r_{+}^{2} + a^{2})}{\Xi\rho_{+}^{2}}\sin^{2}\theta d\wh{\phi}\r)^{2} \\
        & \qquad + \frac{\rho_{+}^{2}}{(r - r_{+})\Delta_{+}'}dr^{2} + \frac{\Delta_{\theta}(r_{+}^{2} + a^{2})^{2}\sin^{2}\theta}{\rho_{+}^{2}\Xi^{2}}d\wh{\phi}^{2}.
    \end{split}
\end{align}
The inverse temperature is easily read off in the $(\tau,\wh{\phi})$ coordinates
\begin{gather}
    \label{Kerr temperature 1}
    \beta = \frac{4\pi(r_{+}^{2} + a^{2})}{\Delta_{+}'} = \frac{2\pi(r_{+}^{2} + a^{2})}{r_{+}(2r_{+}^{2} + a^{2} + 1)}.
\end{gather}
Now, along with the $2\pi$ periodicity in $\phi$, we have the following periodic identifications in the $(\tau,\wh{\phi})$ coordinates
\begin{gather}
    \label{Kerr periodicity 1}
    (\tau,\wh{\phi}) \sim (\tau + \beta,\wh{\phi}) \sim (\tau,\wh{\phi} + 2\pi).
\end{gather}
In the original $(\tau,\phi)$ coordinates, the periodicity is stated as
\begin{gather}
    \label{Kerr periodicity 2}
    (\tau,\phi) \sim \l(\tau + \beta,\phi + i\beta\frac{a\Xi}{r_{+}^{2} + a^{2}}\r) \sim (\tau,\phi + 2\pi).
\end{gather}

It is tempting to identify the shift of $\phi$ in the first identification in \eqref{Kerr periodicity 2} with $i\beta\O$. But this is not correct. As explained in \cite{Gibbons:2004ai}, in the $\phi$ coordinate the asymptotic boundary of AdS itself rotates. To see this, we analyze the metric in \eqref{Kerr metric 4} at large $r$
\begin{align}
    \notag
    ds^{2} & = r^{2}\l(d\tau - i\frac{a}{\Xi}\sin^{2}\theta d\phi\r)^{2} + \frac{dr^{2}}{r^{2}} + \frac{r^{2}\Delta_{\theta}\sin^{2}\theta}{\Xi^{2}}d\phi^{2} \\
    \label{Kerr metric 8}
    \Rightarrow ds^{2} & = \frac{r^{2}\Delta_{\theta}}{\Xi}d\tau^{2} + \frac{dr^{2}}{r^{2}} + \frac{r^{2}\sin^{2}\theta}{\Xi}(d\phi - iad\tau)^{2}.
\end{align}
We note that the asymptotic metric contains a $d\phi d\tau$ term, indicating the fact that the boundary is rotating in these coordinates. The correct variable with respect to which we define the rotation 
of the black hole is given by
\begin{gather}
    \label{Coordinate change 1}
    \wt{\phi} = \phi - ia\tau.
\end{gather}
The periodicity \eqref{Kerr periodicity 2} can be restated in the $(\tau,\wt{\phi})$ coordinates as
\begin{gather}
    \label{Kerr periodicity 3}
    (\tau,\wt{\phi}) \sim (\tau + \beta,\wt{\phi} + i\beta\O) \sim (\tau,\wt{\phi} + 2\pi),
\end{gather}
where we have identified the angular velocity
\begin{gather}
    \label{Kerr angular velocity 1}
    \O = \frac{a\Xi}{r_{+}^{2} + a^{2}} + a = \frac{a(r_{+}^{2} + 1)}{r_{+}^{2} + a^{2}}.
\end{gather}


\subsection{Euclidean path integral and on-shell action}
Having obtained the expressions for the inverse temperature and angular velocity, we now compute the partition function defined by the following Euclidean path integral
\begin{gather}
    \label{Kerr partition function 1}
    \mcz = \int{[Dg]}\exp(-I).
\end{gather}
The Gibbons-Hawking prescription \cite{Gibbons:1976ue} tells us to integrate over Euclidean geometries that have the same induced metric at the asymptotic boundary as the one obtained from the rotating black hole metric in \eqref{Kerr metric 4}. We cannot simply define $I$ as the usual gravity action because of the infinite volume divergence of AdS space. We regulate this divergence by taking the asymptotic boundary at a cutoff radius $r = R_{c}$ and defining $I$ as the background subtracted action\footnote{For AdS spaces, another way to regularize the volume divergence is by using boundary counter-terms in the action \cite{Skenderis:2002wp}. In this approach, the renormalized action differs from the background subtraction one by a constant corresponding to the Casimir energy (see \cite{Gibbons:2005jd,Awad:2007me}).}
\begin{gather}
    \label{Background subtracted action 1}
    I = S_{g} - S_{\rm AdS}.
\end{gather}
Here, $S_{g}$ is the usual Euclidean gravity action
\begin{gather}
    \label{Background subtracted action 2}
    S_{g} = -\frac{1}{16\pi G}\int{d^{5}x}\sqrt{g}(R + 12)
\end{gather}
and $S_{\rm AdS}$ is its value for the thermal AdS solution satisfying the correct boundary conditions. Note that in defining \eqref{Background subtracted action 2}, we have only kept the Einstein-Hilbert term along with the negative cosmological constant. In general, we also need the GHY boundary term but it is easy to check that it does not contribute to the background subtracted action. This is because the leading contribution of the boundary term is same for asymptotically AdS solutions as $R_{c}$ is taken to be large. \\

Let us now evaluate the on-shell action for the rotating black hole solution in \eqref{Kerr metric 4}. Since there is no matter source, the Einstein equations fix the curvature scalar to $R = -20$. The action is then evaluated as
\begin{align}
    \notag
    S_{\rm BH} & = -\frac{1}{16\pi G}\int{d\O_{3}}\int_{0}^{\beta}{d\tau}\int_{r_{+}}^{R_{c}}{dr}\frac{(r^{2} + a^{2}\cos^{2}\theta)r}{\Xi}(-20 + 12), \\
    \label{Kerr action 1}
    \Rightarrow S_{\rm BH} & = \frac{\pi\beta}{4\Xi G}(R_{c}^{4} - r_{+}^{4} + a^{2}R_{c}^{2} - a^{2}r_{+}^{2}).
\end{align}
Note that we have used Hopf coordinates for the integration on $S^{3}$ so that $\int{d\O_{3}} = \int_{0}^{2\pi}{d\phi}\int_{0}^{2\pi}{d\psi}\int_{0}^{\frac{\pi}{2}}{d\theta}\sin\theta\cos\theta$. \\

To compute the background subtracted action for the black hole we need the on-shell action for the appropriate AdS solution.
The solution with the correct induced metric at the asymptotic boundary can be obtained as follows.
First consider the Euclidean AdS metric in global coordinates $(\tau,y,\vartheta,\Phi,\psi)$ with radial coordinate $y$ and the boundary at $y \to \infty$.
A transformation to coordinates $(\tau,r,\theta,\phi,\psi)$ given by
\begin{align}
    \label{Kerr coordinate transformation 1}
    y^{2}\sin^{2}\vartheta & = \frac{(r^{2} + a^{2})\sin^{2}\theta}{\Xi}, \\
    \label{Kerr coordinate transformation 2}
    y^{2}\cos^{2}\vartheta & = r^{2}\cos^{2}\theta, \\
    \label{Kerr coordinate transformation 3}
    \Phi & = \phi - ia\tau,
\end{align}
brings the AdS metric to the form \eqref{Kerr metric 4} with the function $\Delta_{r}(r)$ replaced by
\begin{gather}
    \label{Kerr metric 9}
    \Delta_{r}^{\rm AdS}(r) = (r^{2} + a^{2})(r^{2} + 1).
\end{gather}
Note that the $\Phi$ coordinate is simply what we defined as $\wt{\phi}$ in \eqref{Coordinate change 1}. \\

We compute the AdS action directly in the $(\tau,r,\theta,\phi,\psi)$ coordinates.
Setting $y = 0$ in \eqref{Kerr coordinate transformation 1}--\eqref{Kerr coordinate transformation 2} gives the lower limit of the $r$--integration as $r = 0$.
The upper limit is simply the cutoff radius $r = R_{c}$.
The period of $\tau$ integration in the AdS metric is taken as $\beta'$.
Matching the induced metric at the cutoff surface for the AdS and black hole solutions relates $\beta$ and $\beta'$ as follows
\begin{gather}
    \label{Kerr temperature 2}
    \sqrt{\Delta_{r}^{\rm AdS}(R_{c})}\beta' = \sqrt{\Delta_{r}(R_{c})}\beta.
\end{gather}
For large $R_{c}$, this simplifies to
\begin{gather}
    \label{Kerr temperature 3}
    \beta' = \beta - \frac{\beta}{2R_{c}^{4}}(r_{+}^{2} + a^{2})(r_{+}^{2} + 1) + O\l(\frac{1}{R_{c}^{5}}\r).
\end{gather}
Putting everything together, the action for the AdS solution reads
\begin{align}
    \notag
    S_{\rm AdS} & = -\frac{1}{16\pi G}\int{d\O_{3}}\int_{0}^{\beta'}{d\tau}\int_{0}^{R_{c}}{dr}\frac{(r^{2} + a^{2}\cos^{2}\theta)r}{\Xi}(-20 + 12), \\
    \label{Kerr action 2}
    \Rightarrow S_{\rm AdS} & = \frac{\pi\beta'}{4\Xi G}(R_{c}^{4} + a^{2}R_{c}^{2}).
\end{align}
Finally, to get the background subtracted action for the rotating black hole, we subtract \eqref{Kerr action 2} from \eqref{Kerr action 1}, use the relation \eqref{Kerr temperature 3} and then take the large $R_{c}$ limit
\begin{gather}
    \label{Kerr action 3}
    I_{\rm BH} = \lim_{R_{c} \to \infty}(S_{\rm BH} - S_{\rm AdS}) = \frac{\pi\beta}{8\Xi G}(r_{+}^{2} + a^{2})(1 - r_{+}^{2}).
\end{gather}

In the following sections, we will use the background subtracted action \eqref{Kerr action 3}.
But first, let us comment on a subtlety.
The spatial slices at constant $y$ are spheres while the slices at constant $r$ are spheroids.
In particular, a fixed $y$ surface is described by a complicated relation between $r$ and the angular variables.
To study black hole thermodynamics, it is natural to define the CFT on a non-rotating sphere at large $y$ instead of a surface at large $r$.
Despite this, taking the cutoff surface in the $r$ variable works because at large radius the spheroids are almost spheres and the difference between the two is not seen in background subtraction.
But care is required when using counter-term regularization.
The finite contribution of holographic counter-terms to the on-shell action gives the Casimir energy of the vacuum, which differs between the two choices of asymptotic boundaries \cite{Awad:2007me}.
At large $y$, the renormalized action differs from the background subtracted one by the constant $\frac{3\pi\beta}{32G}$.
In contrast, the large $r$ surface is rotating and the induced metric depends on the parameter $a$.
In turn, the Casimir energy also depends on $a$.


\subsection{Thermodynamics and Hawking-Page phase transition}
\label{section: Thermodynamics and Hawking-Page}
In the ensemble we are considering, we fix the inverse temperature, $\beta$, and rotation, $\O$, by specifying the boundary conditions for the path integral.
From the dual CFT, we know that the partition function makes sense for $\abs{\O} < 1$ for all values of $\beta > 0$.\footnote{For $\abs{\O} \geq 1$, the Hilbert space trace diverges.}
In the AdS bulk, the rotating thermal AdS solution also exists for these values of $\beta$ and $\O$.
Additionally, for small enough $\beta$, we also get two quasi-Euclidean black hole saddles.
These are usually referred to as small and large black holes.
For now, we restrict our attention to these three solutions.
Other (possibly complex) saddles will be discussed in detail in section \ref{section: complex Kerr-AdS solutions}. \\

To recover the thermodynamics of the rotating black holes, we can use the on-shell action computed in \eqref{Kerr action 3}. We write its contribution to the partition function as
\begin{gather}
    \label{BH partition function 1}
    \mcz_{\rm BH} = \exp(-I_{\rm BH}) \times (\cdots),
\end{gather}
where $\cdots$ denotes perturbative loop corrections around the saddle.
Taking appropriate derivatives of $\log\mcz_{\rm BH}$ with respect to $\beta$ and $\O$ reproduces the correct leading order expressions for $E, J$ and $S$
\begin{align}
    \label{Kerr thermodynamics 1}
    E & = \l(-\frac{\p}{\p\beta}+ \frac{\O}{\beta}\frac{\p}{\p\O}\r)\log\mcz_{\rm BH} = \frac{\pi(\Xi + 2)}{8\Xi^{2}G}(r_{+}^{2} + 1)(r_{+}^{2} + a^{2}), \\
    \label{Kerr thermodynamics 2}
    J & = \frac{1}{\beta}\frac{\p}{\p\O}\log\mcz_{\rm BH} = \frac{\pi a}{4\Xi^{2}G}(r_{+}^{2} + 1)(r_{+}^{2} + a^{2}), \\
    \label{Kerr thermodynamics 3}
    S & = \l(-\beta\frac{\p}{\p\beta} + 1\r)\log\mcz_{\rm BH} = \frac{\pi^{2}r_{+}}{2\Xi G}(r_{+}^{2} + a^{2}).
\end{align}
The above expressions hold for both the small and large Kerr-AdS black holes, as well as the complex black holes we will discuss later.
Notably, all extensive quantities for the black hole go as $\frac{1}{G}$.
In particular, the entropy matches with Bekenstein-Hawking formula, $S = \frac{{\rm Area}}{4G}$.
 \\

For the region in $\beta$--$\O$ space where a Kerr-AdS saddle dominates the full path integral, we say that the bulk is in the black hole phase.
In contrast, if the rotating thermal AdS saddle dominates, we call it the thermal phase.
From the perspective of the dual CFT, these are the deconfined and confined phases, respectively.
One way to see this is that the entropy in the deconfined phase goes as $N^{2} \sim \frac{1}{G}$. \\

\begin{figure}[!ht]
    \centering
    \includegraphics[width=0.45\linewidth]{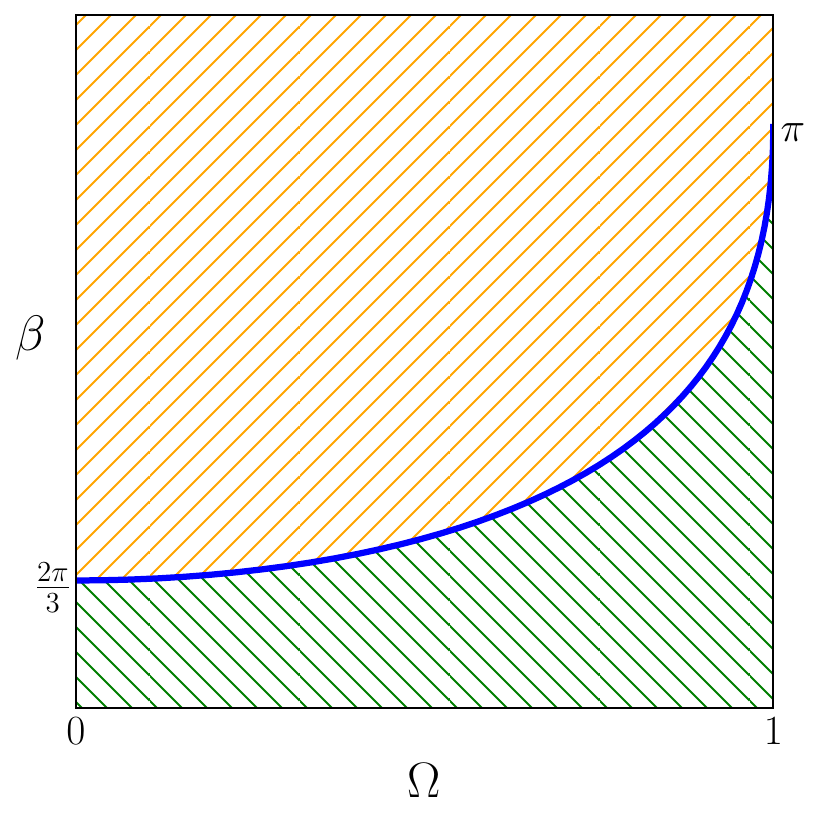}
    \caption{The phase diagram of AdS$_{5}$ in the $\beta$--$\O$ space.
    We only show the region $0 \leq \O < 1$ since the plot for negative values of $\O$ simply mirrors this.
    The black hole phase is indicated by the green shaded region while the thermal phase is the complementary region shaded in orange.
    The Hawking-Page transition curve (blue) separates the two phases.}
    \label{fig: Phase diagram}
\end{figure}

In the situation where both a black hole saddle and the thermal saddle contribute, the one that dominates is decided by which has a smaller on-shell action.
We have already normalized the action in a way that it vanishes for the thermal solution.
So, the black holes dominates for those values of $\beta$ and $\O$ for which $I_{\rm BH} < 0$.
From equation \eqref{Kerr action 3}, we see that this condition is met when $r_{+} > 1$.
Conversely, the thermal solution dominates when $r_{+} < 1$.
Thus, at $r_{+} = 1$ we have a first order phase transition between the black hole and thermal phases.
We interpret this as a confinement-deconfinement phase transition in the dual gauge theory \cite{Witten:1998zw}.
Substituting $r_{+} = 1$ in the thermodynamics relations \eqref{Kerr temperature 1} and \eqref{Kerr angular velocity 1} gives the following Hawking-Page \textit{transition curve}
\begin{gather}
    \label{HP transition 1}
    \beta_{\rm HP}(\O) = \frac{2\pi(2 - \sqrt{1 - \O^{2}})}{3 + \O^{2}}.
\end{gather}
We see that the Hawking-Page (inverse) temperature nicely interpolates between the Schwarzschild value, $\beta_{\rm HP} = \frac{2\pi}{3}$ at $\O = 0$, and $\beta_{\rm HP} = \pi$ at $\O = 1$.
The phase structure of AdS spacetime at finite temperature and rotation is shown in figure \ref{fig: Phase diagram}.


\section{Complex Kerr-AdS Solutions}
\label{section: complex Kerr-AdS solutions}
We now come to the main point of this paper.
In the previous section, we derived the thermodynamic features of AdS spacetime by only considering quasi-Euclidean black holes and the thermal AdS solution.
This would be fine if these were the only saddles of the path integral.
But this is not the case.
We now show that the small and large Kerr-AdS black holes exist as quasi-Euclidean solutions only in part of the $\beta$--$\O$ space.
In the remaining region, they naturally continue over to complex saddles of the path integral.
In effect, this is a generalization to the case with rotation of the observation made in \cite{Mahajan:2025bzo} for complex Schwarzschild black holes.
But, apart from the small and large black holes, we find an additional black hole saddle at nonzero values of $\O$ (for all $\beta$). We call this a \textit{spurious black hole} owing to the fact that it does not have an analogue in the spherically symmetric case. \\

Let us collect the thermodynamic relations \eqref{Kerr temperature 1} and \eqref{Kerr angular velocity 1} here
\begin{align}
    \label{Inverting relations 1}
    \beta & = \frac{2\pi(r_{+}^{2} + a^{2})}{r_{+}(2r_{+}^{2} + a^{2} + 1)}, \\
    \label{Inverting relations 2}
    \O & = \frac{a(r_{+}^{2} + 1)}{r_{+}^{2} + a^{2}}.
\end{align}
It would be useful if we can invert these relations and write $r_{+}$ and $a$ directly in terms of $\beta$ and $\O$.
In the AdS-Schwarzschild case, i.e. setting $a = 0$, inverting the relation \eqref{Inverting relations 1} gives two solutions for the horizon radius
\begin{gather}
    \label{Schwarzschild horizon radius 1}
    r_{+}^{\pm} = \frac{\pi \pm \sqrt{\pi^{2} - 2\beta^{2}}}{2\beta}.
\end{gather}
These are the small ($r_{+}^{-}$) and large ($r_{+}^{+}$) spherical black holes.
When we turn on a finite $\O$, we obtain the rotating counterparts of these solutions.
The expressions for general $\O$ look complicated, but for $\O \ll 1$ we can expand around the spherical case to write
\begin{align}
    \label{Inverting relations 3}
    r_{+} & = r_{0} + \frac{r_{0}^{3}}{2r_{0}^{4} + r_{0}^{2} - 1}\O^{2} + O(\O^{4}), \\
    \label{Inverting relations 4}
    a & = \frac{r_{0}^{2}}{r_{0}^{2} + 1}\O + \frac{r_{0}^{4}(2r_{0}^{2} + 1)}{(r_{0}^{2} + 1)^{3}(2r_{0}^{2} - 1)}\O^{3} + O(\O^{5}).
\end{align}
Here, $r_{0}$ takes either of the the values in \eqref{Schwarzschild horizon radius 1}.
The above expressions describe the small ($r_{0} = r_{+}^{-}$) and large ($r_{0} = r_{+}^{+}$) Kerr-AdS black holes. \\

\begin{figure}[!ht]
    \centering
    \includegraphics[width=0.45\linewidth]{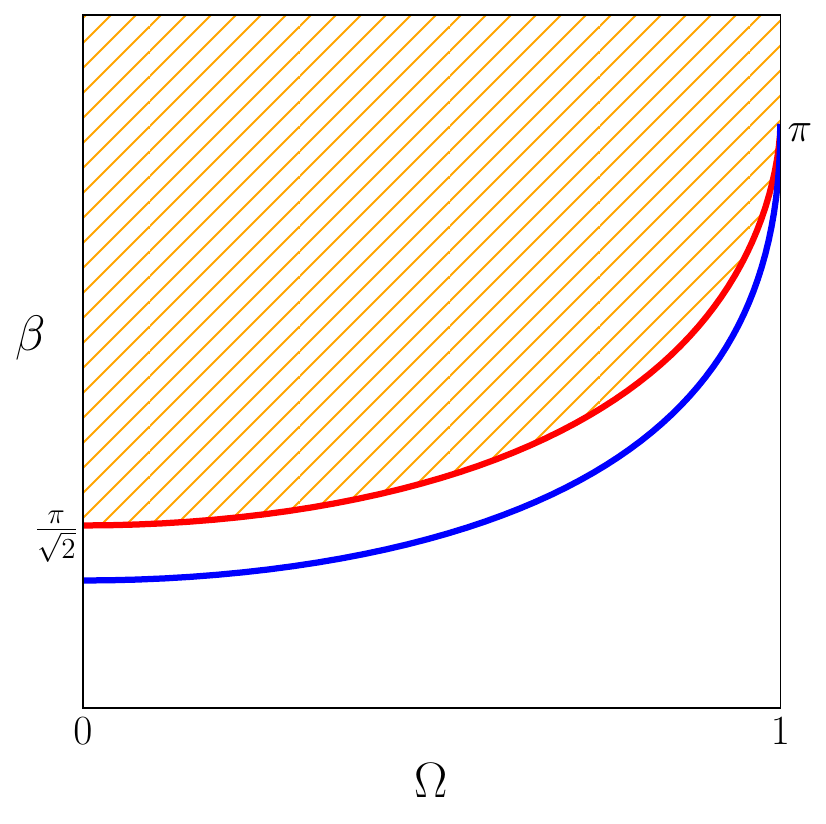}
    \caption{We plot the curve $\beta_{\rm max}(\O)$ (red) in $\beta$--$\O$ space.
    The Hawking-Page transition curve (blue) is shown for comparison.
    The orange shaded part corresponds to $\beta > \beta_{\rm max}(\O)$, the region for which rotating black holes are complex.
    So, black holes become subdominant saddles before they become complex.}
    \label{fig: Coalescing saddles}
\end{figure}

It is clear that for small enough values of $\beta$, the relation \eqref{Schwarzschild horizon radius 1} returns real values of $r_{+}^{\pm}$ so that the AdS-Schwarzschild black holes are real.
But, as pointed out in \cite{Mahajan:2025bzo}, as $\beta$ is increased all the way to $\beta_{\rm max} = \frac{\pi}{\sqrt{2}}$, the two solutions coalesce with $r_{+}^{\pm} = \frac{1}{\sqrt{2}}$.
A further increase in $\beta$ leads to complex solutions which were dubbed as complex AdS-Schwarzschild black holes.
We now make an analogous observation in the present case of finite rotation.
At fixed $\O$, there exists a corresponding value $\beta_{\rm max}(\O)$ at which the two rotating black holes appear as a repeated solution of equations \eqref{Inverting relations 1} and \eqref{Inverting relations 2}.
The curve $\beta_{\rm max}(\O)$ is determined by demanding the proportionality of tangent vectors at the solution 
\begin{gather}
    \label{Coalescing saddles 1}
    \l(\frac{\p\beta}{\p r_{+}},\frac{\p\beta}{\p a}\r) \propto \l(\frac{\p\O}{\p r_{+}},\frac{\p\O}{\p a}\r).
\end{gather}
This condition leads to the following parametric form for $\beta_{\rm max}(\O)$
\begin{align}
    \label{Coalescing saddles 2}
    \beta_{\rm max} & = \frac{\pi}{\sqrt{2}}\frac{3a^{2} + 1}{(a^{2} + 1)^{\frac{3}{2}}}, \\
    \label{Coalescing saddles 3}
    \O & = \frac{a(a^{2} + 3)}{3a^{2} + 1},
\end{align}
where we take $\abs{a} < 1$ so that $\abs{\O} < 1$.\footnote{Explicit expression for $\beta_{\rm max}(\O)$ can be obtained by eliminating $a$, but it is not very enlightening.}
So, for $0 < \beta < \beta_{\rm max}({\O})$ we get quasi-Euclidean black holes while for $\beta > \beta_{\rm max}(\O)$ the black holes are genuinely complex.
The curve $\beta_{\rm max}(\O)$ is shown in figure \ref{fig: Coalescing saddles} alongside the transition curve $\beta_{\rm HP}(\O)$.
~\\

Based on the analysis of section \ref{section: Thermodynamics and Hawking-Page} and the discussion above, we can form the following conclusions.
At a fixed value of the rotation and small enough inverse temperature, the large rotating black hole is a quasi-Euclidean solution that dominates over rotating thermal AdS.
As inverse temperature is increased, the black holes first become subdominant and then complex.
It is easy to see that this is actually the situation with AdS black holes in all spacetime dimensions four and higher. \\

However, the complex black holes are \textit{not subdominant} at all the temperatures.
In five spacetime dimensions (and higher), as $\beta$ is increased beyond $\beta_{\rm max}$, the real part of the on-shell action again becomes negative, thus implying that the complex black holes can come back to dominate over the thermal solution.
In the extreme case, $\beta \to \infty$ with $\O$ fixed, equations \eqref{Inverting relations 1}--\eqref{Inverting relations 2} return
\begin{gather}
    \label{Extreme case 1}
    r_{+} = \pm i\sqrt{\frac{1 + \O^{2}}{2}}, \quad a = -\O.
\end{gather}
The on-shell action \eqref{Kerr action 3} at these values of the parameters evaluates to
\begin{gather}
    \label{Extreme case 2}
    I_{\rm BH} = -\frac{\pi\beta(3 + \O^{2})}{32G}.
\end{gather}
This means that if the complex black holes do contribute to the AdS thermodynamics, then apart from the Hawking-Page phase transition there is another phase transition (at a lower temperature) back to the black hole phase.
Again, by AdS/CFT, this implies an additional confinement-deconfinement phase transition in the dual gauge theory.
But, this goes against the general expectation that at low enough temperatures a large--$N$ gauge theory is confined \cite{Aharony:2003sx}. \\

We point out another related issue.
It turns out that at a given value of the inverse temperature and rotation, there is another black hole solution apart from the two (quasi-Euclidean or complex) Kerr-AdS ones.
This additional solution does not admit a nice $\O \to 0$ limit and hence does not have an analogue in the spherically symmetric case.
For this reason, we call it a \textit{spurious black hole}.
To construct this black hole, we look for a solution of the relations \eqref{Inverting relations 1}--\eqref{Inverting relations 2} that has a singular $\O \to 0$ limit.
We can write the solution as an expansion in $\O$ as follows
\begin{align}
    \label{Inverting relations 5}
    r_{+} & = r_{0} - \frac{r_{0}}{r_{0}^{2} + 1}\O^{2} + O(\O^{4}), \\
    \label{Inverting relations 6}
    a & = (r_{0}^{2} + 1)\frac{1}{\O} - \frac{3r_{0}^{2}}{r_{0}^{2} + 1}\O + O(\O^{3}),
\end{align}
where $r_{0} \equiv \frac{2\pi}{\beta}$. With these values of the black hole parameters, the Kerr-AdS metric \eqref{Kerr metric 4} is not quasi-Euclidean.
Indeed, the ``Lorentzian" metric \eqref{Kerr metric 1} for this solution does not have Lorentzian signature.
Rather, it describes a signature change metric. \\

Despite the fact the spurious black hole does not have a well-defined limit to the spherically symmetric case, it is a legitimate saddle for the path integral that computes the partition function at nonzero rotation.
Plugging in the values \eqref{Inverting relations 5}-\eqref{Inverting relations 6} into \eqref{Kerr action 3} gives the following expansion for the on-shell action
\begin{gather}
    \label{SpBH action 1}
    I_{\rm BH} = \frac{\pi\beta(r_{0}^{2} - 1)}{8G} - \frac{\pi\beta}{8G}\O^{2} + O(\O^{4}).
\end{gather}
Curiously, the action is finite in the limit $\O \to 0$.
Furthermore, from the leading term in \eqref{SpBH action 1}, we note that the on-shell action is negative for $r_{0} < 1$, or equivalently, for $\beta > 2\pi$.
So, at small enough temperatures, even the spurious black hole is dominant over the thermal AdS saddle.
In fact, comparing \eqref{SpBH action 1} against \eqref{Extreme case 2} in the extreme case $\beta \to \infty$, we see that it is dominant over the complex Kerr-AdS solutions as well.
This means that if the spurious black hole also contributes to the partition function, then there is a \textit{third phase transition}, this time between two black hole phases.
Again, this is an unexpected feature from the point of view of the dual gauge theory. \\

At small $\abs{\O}$, it is easy to verify that the only solutions to the relations \eqref{Inverting relations 1}--\eqref{Inverting relations 2} are the small and large Kerr-AdS black holes, as well as the spurious black hole.
Since the number of solutions for generic values of $\beta$ and $\O$ remains the same, we have not missed any solution. \\

To remedy the puzzle with the additional phase transitions, we proceed along the lines of \cite{Mahajan:2025bzo}.
In section \ref{section: minisuperpsace}, we use an appropriate mini-superspace approximation for the original path integral and analyze the residual finite-dimensional integral via Picard-Lefschetz theory.


\section{Mini-superspace Analysis}
\label{section: minisuperpsace}
We want to carefully understand which saddles contribute to the partition function as the boundary conditions (values of $\beta$ and $\O$) of the path integral \eqref{Kerr partition function 1} are varied.
A direct approach towards understanding the full infinite-dimensional integral is not available at present.
Instead, we assume that there exists a consistent truncation of the path integral to a finite-dimensional integral that can still capture some of the essential features, such as the information of which saddles contribute.
The residual integral is called a mini-superspace approximation, and it can be analyzed using steepest descent, or more generally, Picard-Lefschetz theory. \\

For the case in hand, since we are asymptotically fixing $\beta$ and $\O$, the energy ($E$) and angular momentum ($J$) are integration variables in the path integral.
We can then write down a mini-superspace approximation assuming that the remaining (infinitely many) variables of the path integral have already been integrated over, leaving behind the $E$ and $J$ integrations \cite{Marolf:2022ybi}.
Alternatively, we can take $r_{+}$ and $a$ as a convenient choice of integration variables and write the partition function as
\begin{gather}
    \label{Minisuperspace 1}
    \mcz(\beta,\O) = \int_{0}^{\infty}{dr_{+}}\int_{-1}^{1}{da}\,\exp[S - \beta(E - \O J)],
\end{gather}
where the quantities $E, J$ and $S$ were obtained as functions of $r_{+}$ and $a$ in equations \eqref{Kerr thermodynamics 1}--\eqref{Kerr thermodynamics 3}.
Note that the domain of integration in the $r_{+},a$ variables corresponds to $0 \leq \abs{J} < E < \infty$.
In writing \eqref{Minisuperspace 1}, we have only kept functions which go as $\frac{1}{G}$ in the exponent.
This is because we are keeping track of which saddles contribute but not of the one-loop (and higher) corrections. \\

By our very construction of the integrand in \eqref{Minisuperspace 1}, rotating thermal AdS and all the rotating black hole solutions (quasi-Euclidean, complex, and spurious) satisfying the thermodynamics relations \eqref{Inverting relations 1} and \eqref{Inverting relations 2} appear as saddle points of the exponent.
To answer the question of which saddles contribute at a given $\beta$ and $\O$, we need to decompose the integration region of \eqref{Minisuperspace 1} into a collection of Lefschetz thimbles.
Explicitly, we have
\begin{gather}
    \label{Decomposition 1}
    \mcc = \sum_{\sigma \in \,\text{saddles}}n_{\sigma}\mcj_{\sigma},
\end{gather}
where $\mcc$ denotes the integration region while $\mcj_{\sigma}$ corresponds to the Lefschetz thimble of saddle $\sigma$ with a chosen orientation.
At each $\beta$ and $\O$, we want to know the values of $n_{\sigma}$ to decide which saddles contribute to the integral.
We know that as the parameters are varied continuously, the saddles and the thimbles also change continuously so that the numbers $n_{\sigma}$ remain constant.
These numbers only change when we hit a Stokes surface, where the thimbles also jump discontinuously.
From the earlier discussion in section \ref{section: complex Kerr-AdS solutions}, it is clear that the curve $\beta = \beta_{\rm max}(\O)$ for $\abs{\O} < 1$, is (part of) the Stokes surface.
So, we expect that the decomposition \eqref{Decomposition 1} takes one form for $\beta > \beta_{\rm max}(\O)$ and another for $\beta < \beta_{\rm max}(\O)$.
We will study the two cases separately. \\

The case $\beta > \beta_{\rm max}(\O)$ is the problematic one where the small and large Kerr-AdS solutions, as well as the spurious black hole, appear as complex saddle points that can dominate over the rotating thermal AdS saddle at large values of $\beta$.
The thermal saddle already lies on the integration region $\mcc$.
We now need to see how each thimble contributes in the decomposition \eqref{Decomposition 1}. \\

Let us briefly discuss how Lefschetz thimbles for each saddle are constructed (see \cite{Witten:2010cx,Dunne:2015eaa} for detailed reviews).
We start by writing the exponent in $\eqref{Minisuperspace 1}$ as
\begin{gather}
    \label{Exponent 1}
    \mci(r_{+},a) = S(r_{+},a) - \beta(E(r_{+},a) - \O J(r_{+},a)).
\end{gather}
A saddle point $p_{\sigma} = (r_{+,\sigma},a_{\sigma})$ is a solution to the equations
\begin{gather}
    \label{Saddle equation 1}
    \frac{\p\mci}{\p r_{+}} = 0, \quad \frac{\p\mci}{\p a} = 0.
\end{gather}
The corresponding thimble is denoted as $\mcj_{\sigma}$ and is defined as the collection of downward flow trajectories emanating from the saddle.
Explicitly, the trajectories $(r_{+}(t),a(t))$ solve the flow equations
\begin{gather}
    \label{Flow equations 1}
    \frac{du^{i}}{dt} = -\frac{\p\Re(\mci)}{\p u^{i}},
\end{gather}
where $u^{i} = (\Re(r_{+}),\Im(r_{+}),\Re(a),\Im(a))$, with the condition $(r_{+}(t),a(t)) = p_{\sigma}$ at $t = -\infty$.
Near the saddle point, the direction of the flow is determined by elements of the Hessian matrix. \\

Substituting in \eqref{Exponent 1} the expressions \eqref{Kerr thermodynamics 1}--\eqref{Kerr thermodynamics 3} for $E, J$ and $S$ gives
\begin{gather}
    \label{Exponent 2}
    \mci(r_{+},a) =\frac{\pi}{8G}\frac{r_{+}^{2} + a^{2}}{(1 - a^{2})^{2}}
    (4\pi r_{+}(1 - a^{2}) - \beta(r_{+}^{2} + 1)(3 - a^{2} - 2a\O)).
\end{gather}
The function is analytic except for poles at $a = \pm 1$. A gradient flow (upward or downward) from a generic point will asymptote towards one of the poles.\footnote{Along a downward (upward) flow the real part of the exponent decreases (increases) indefinitely. Since \eqref{Exponent 2} has a finite limit as $\abs{a} \to \infty$, the only other option for a flow line is to approach a pole.}
The thermal solution ($p_{0} = (0,0)$) automatically satisfies the saddle equations. The corresponding elements of the Hessian matrix are easily computed as
\begin{gather}
    \label{Hessian thermal 1}
    \eval{\frac{\p^{2}\mci}{\p r_{+}^{2}}}_{p_{0}} = -\frac{3\pi}{4G}\beta,
    \quad \eval{\frac{\p^{2}\mci}{\p a^{2}}}_{p_{0}} = -\frac{3\pi}{4G}\beta,
    \quad \eval{\frac{\p^{2}\mci}{\p r_{+}\p a}}_{p_{0}} = 0.
\end{gather}
This tells us that the downward flow trajectories beginning from the thermal saddle move in the $\Re(r_{+})$ and $\Re(a)$ directions.
Since the function $\mci$ is ``real-valued" for real $\beta, \O$ (and $G$), these flow lines lie entirely on the real plane $\Im(r_{+}) = 0, \Im(a) = 0$. \\

\begin{figure}[!ht]
    \centering
    \includegraphics[width=0.65\textwidth]{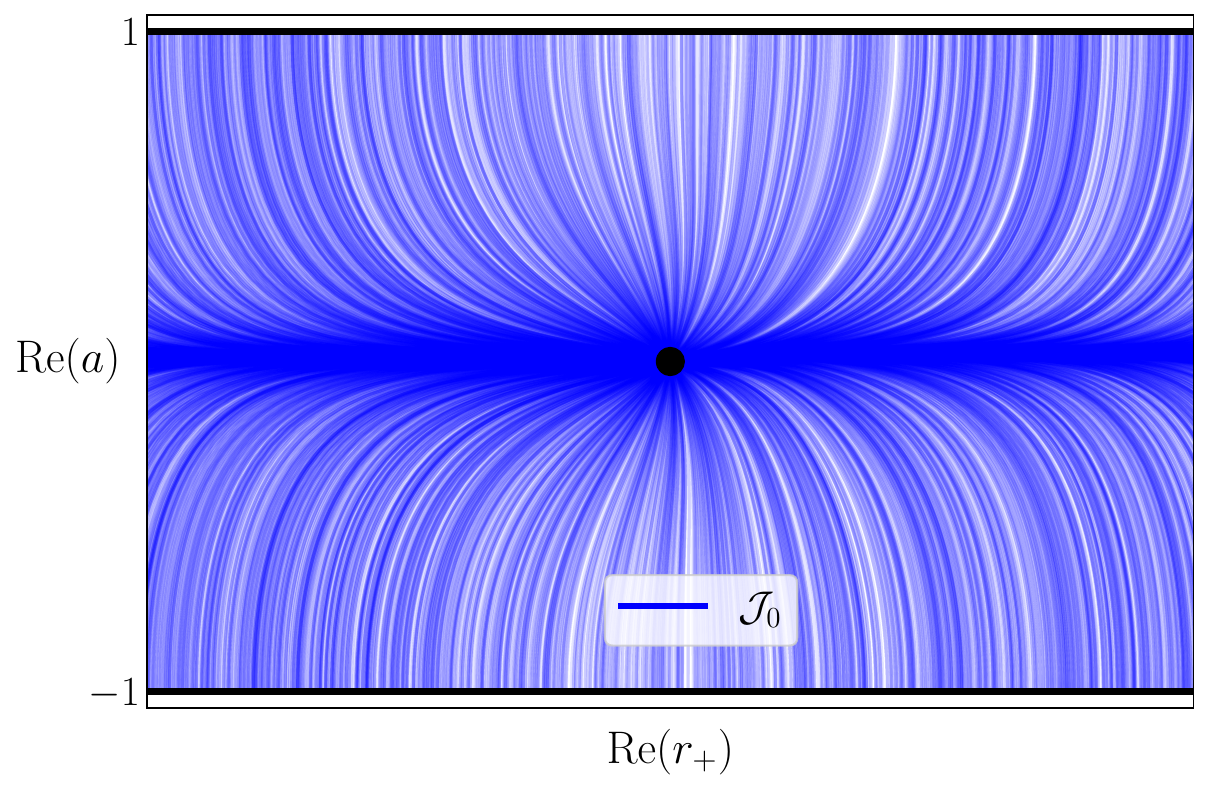}
    \caption{A region on the real plane ($\Im(r_{+}) = 0, \Im(a) = 0$) showing downward flows from the thermal saddle $p_{0}$ (thick black dot) for $\beta = \pi > \beta_{\rm max}(\O)$ and $\O = 0.1$\,.
    The thimble $\mcj_{0}$ is a union of trajectories (blue) that start at $p_{0}$ and asymptote towards $r_{+}  = \pm \infty, a = \pm 1$.}
    \label{fig: Thimble 1}
\end{figure}

In the case $\beta > \beta_{\rm max}(\O)$, the flow lines do not encounter another saddle and end up spanning the region described by $-\infty < r_{+} < \infty, -1 < a < 1$.
The Lefschetz thimble $\mcj_{0}$ is precisely this subspace of the full space.
We show it as a collection of flow lines emanating from $p_{0}$ in figure \ref{fig: Thimble 1}.
Since the original integration region is a part this subspace, the decomposition \eqref{Decomposition 1} into thimbles is
\begin{gather}
    \label{Decomposition 2}
    \mcc = \frac{1}{2}\mcj_{0}.
\end{gather}
This means that for any value of the rotation and for any value of the inverse temperature larger than $\beta_{\rm max}(\O)$, the complex and the spurious black holes of section \ref{section: complex Kerr-AdS solutions} \textit{do not} contribute to the partition function, regardless of the value of their on-shell action.
This resolves the puzzle we raised earlier.

Now, let us consider the case $\beta < \beta_{\rm max}(\O)$, for which the small and large black holes are real saddle points. We indicated earlier that the qualitative behaviour of Lefschetz thimbles changes discontinuously as a parameter moves across a Stokes surface.
This raises the following question: what happens to the decomposition \eqref{Decomposition 1} as the inverse temperature is lowered below $\beta_{\rm max}(\O)$ at a given $\O$.
In other words, how do contributions of the quasi-Euclidean small and large black holes, along with that of thermal AdS, enter the partition function.
Recall that the analogous question in the case without rotation was addressed in \cite{Mahajan:2025bzo}.
It was concluded that only thermal AdS and the large black hole contribute to the partition function. \\

\begin{figure}[!ht]
    \centering
    \includegraphics[width=0.65\textwidth]{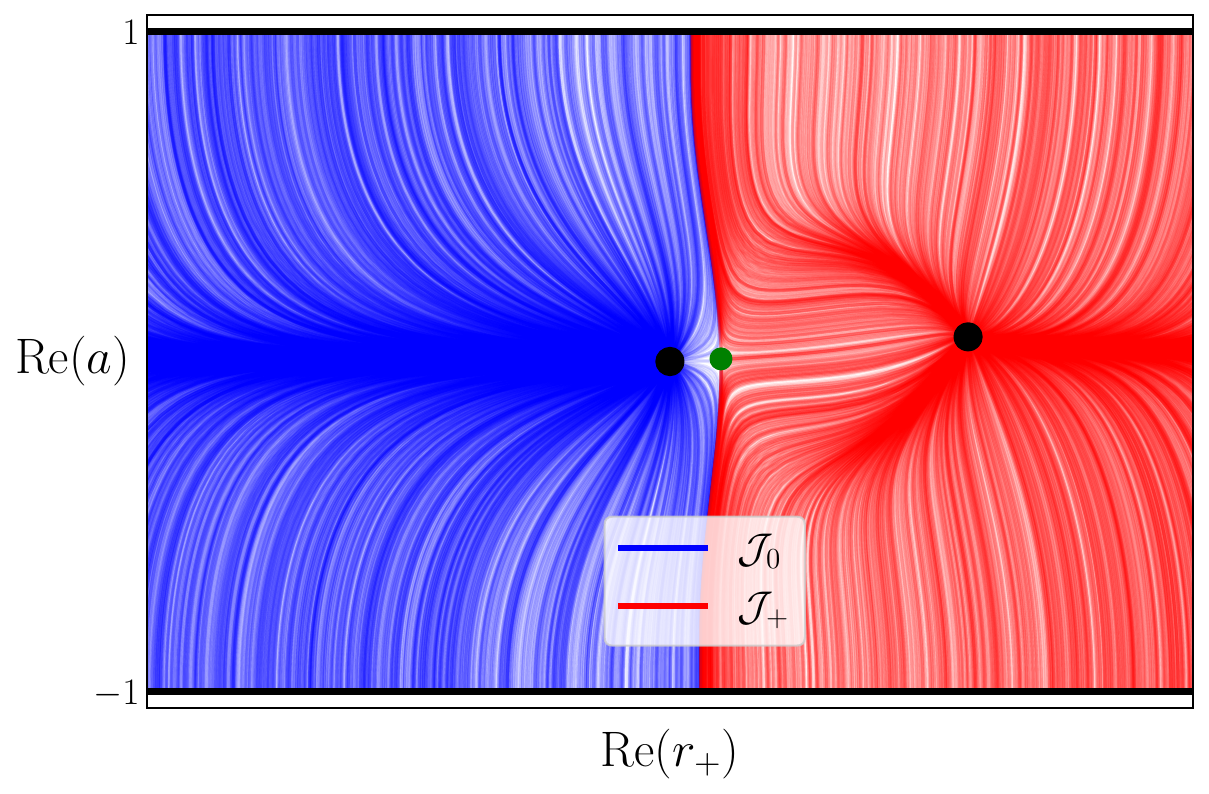}
    \caption{A region on the real plane ($\Im(r_{+}) = 0, \Im(a) = 0$) showing flow lines emanating from the thermal saddle $p_{0}$ and the large black hole saddle $p_{+}$ (thick black dots) for $\beta = \frac{\pi}{2} < \beta_{\rm max}(\O)$ and $\O = 0.1$\,.
    The thimble $\mcj_{0}$ is a union of trajectories (blue) that start at $p_{0}$ and asymptote towards $r_{+} = -\infty, a = \pm 1$, while the thimble $\mcj_{+}$ is a union of trajectories (red) that start at $p_{+}$ and asymptote towards $r_{+} = \infty, a = \pm 1$.
    The small black hole saddle $p_{-}$ is marked by a green dot.
    Flow lines from $p_{-}$ go in the $\Im(r_{+})$ direction.}
    \label{fig: Thimble 2}
\end{figure}

Even in the present scenario, the Hessian elements obtained in \eqref{Hessian thermal 1} for the thermal saddle are negative (for $G > 0$), so the discussion surrounding the equation holds and the downward flows lie on the real plane.
The small ($p_{-}$) and large ($p_{+}$) black hole saddles are also real and lie in the region $-\infty < r_{+} < \infty, -1 < a < 1$, so there can be flow lines between two saddles.
The fact that the imaginary part of the action remains constant along flow lines certainly allows for such flows, since the action is real-valued when the parameters are real.
We show flow lines corresponding to thermal and large black hole saddles in figure \ref{fig: Thimble 2}.
It is clear that there is a downward flow from $p_{0}$ to $p_{-}$ as well as one from $p_{+}$ to $p_{-}$.
This means that the entire parameter space $0< \beta < \beta_{\rm max}(\O)$ for $\abs{\O} < 1$ and real $G$ is part of the Stokes surface.

If we forget for the time being about the fact that we are on a Stokes surface, a direct attempt at decomposing the integration region $\mcc$ in terms of the thimbles shown in figure \ref{fig: Thimble 2} would read as
\begin{gather}
    \label{Decomposition 3}
    \mcc = \frac{1}{2}\mcj_{0} + \mcj_{+}.
\end{gather}
If this is true, we can conclude that the small Kerr-AdS black hole does not contribute to the partition function at fixed $\beta$ and $\O$, even in a suppressed manner. But since some thimbles change discontinuously, a thimble decomposition on the Stokes surface is not directly meaningful. \\

The usual resolution of this problem is by deforming the parameters in a way that we are no longer on the Stokes surface.
We proceed by giving a small imaginary part to $G$ (keeping $\Re(G) > 0$).
Since it appears as an overall factor, the saddles themselves do not change, but the imaginary part of the on-shell action modifies.
In particular, the imaginary part of the on-shell action is unequal for the three saddles meaning that there are no flow lines between the saddles. \\

\begin{figure}[!ht]
    \centering
    \includegraphics[width = 0.49\textwidth]{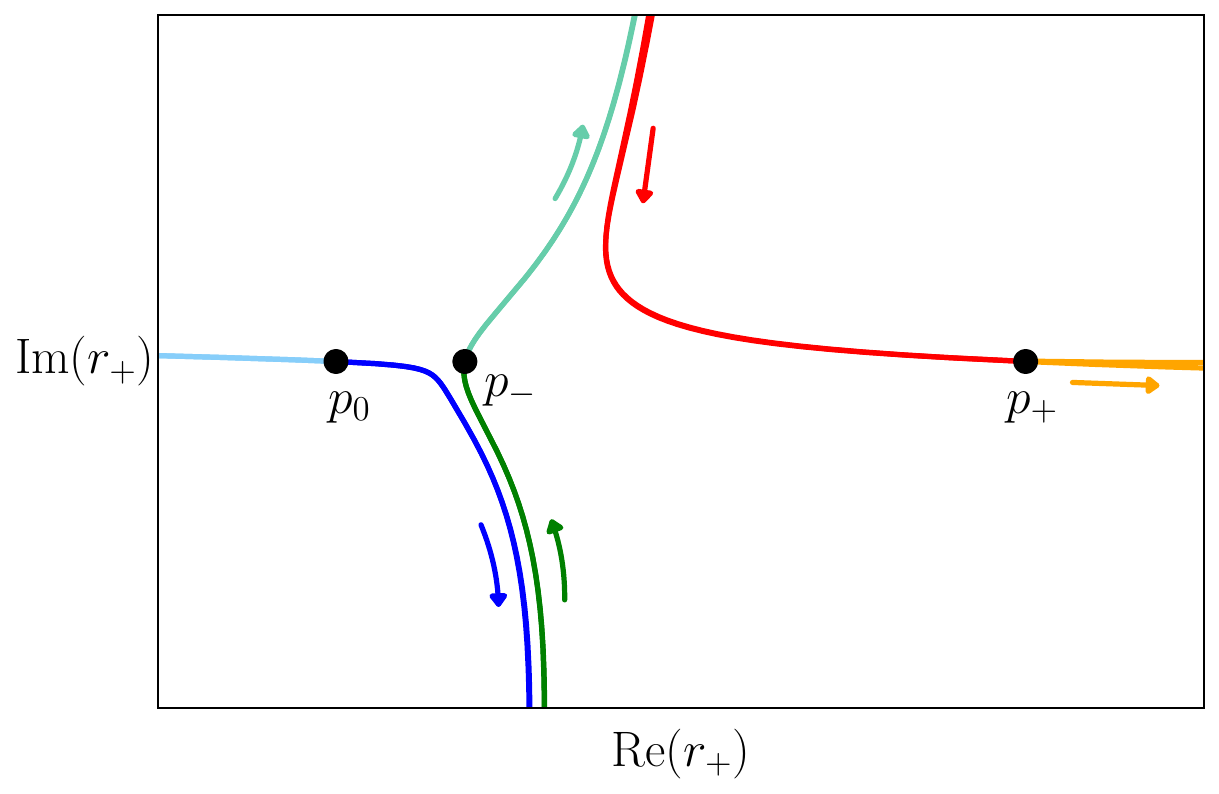}
    \includegraphics[width = 0.49\textwidth]{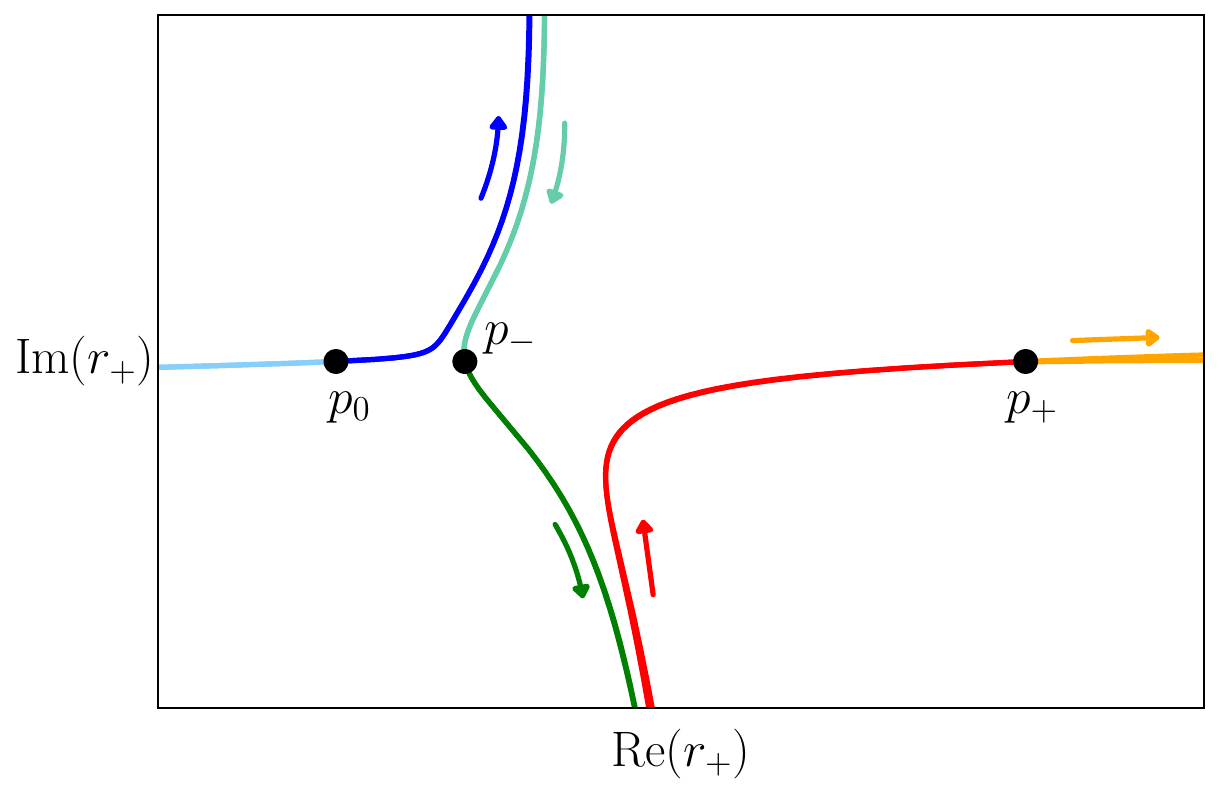}
    \caption{Projection on $r_{+}$--plane of downward flow lines from the three saddles with $\Im(G) < 0$ (left) and $\Im(G) > 0$ (right) in the case $\beta = \frac{\pi}{2} < \beta_{\rm max}(\O)$ and $\O = 0.1$\,.
    Each saddle $p_{\sigma}$ is marked by a thick black dot.
    The flow lines are colour coded according to their asymptotic behaviour.}
    \label{fig: Projection on rplus plane}
\end{figure}

\begin{figure}[!ht]
    \centering
    \includegraphics[width = 0.49\textwidth]{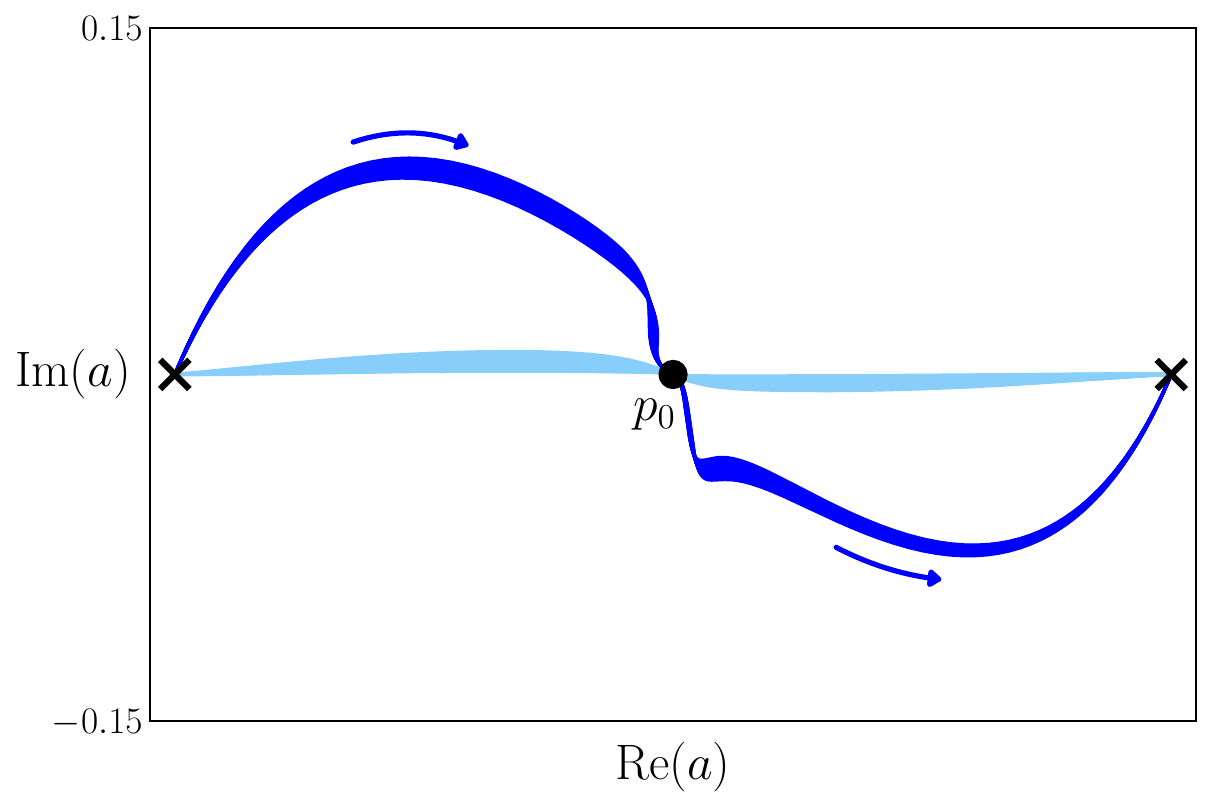}
    \includegraphics[width = 0.49\textwidth]{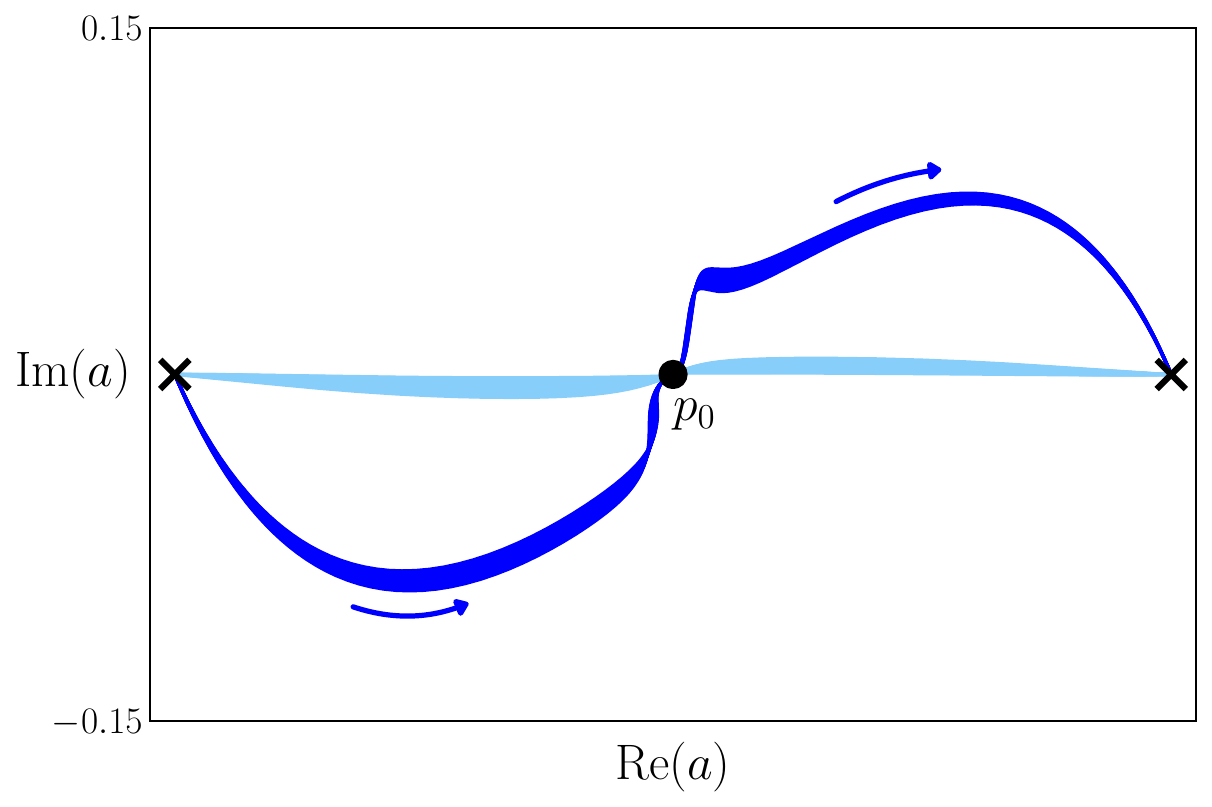}
    \includegraphics[width = 0.49\textwidth]{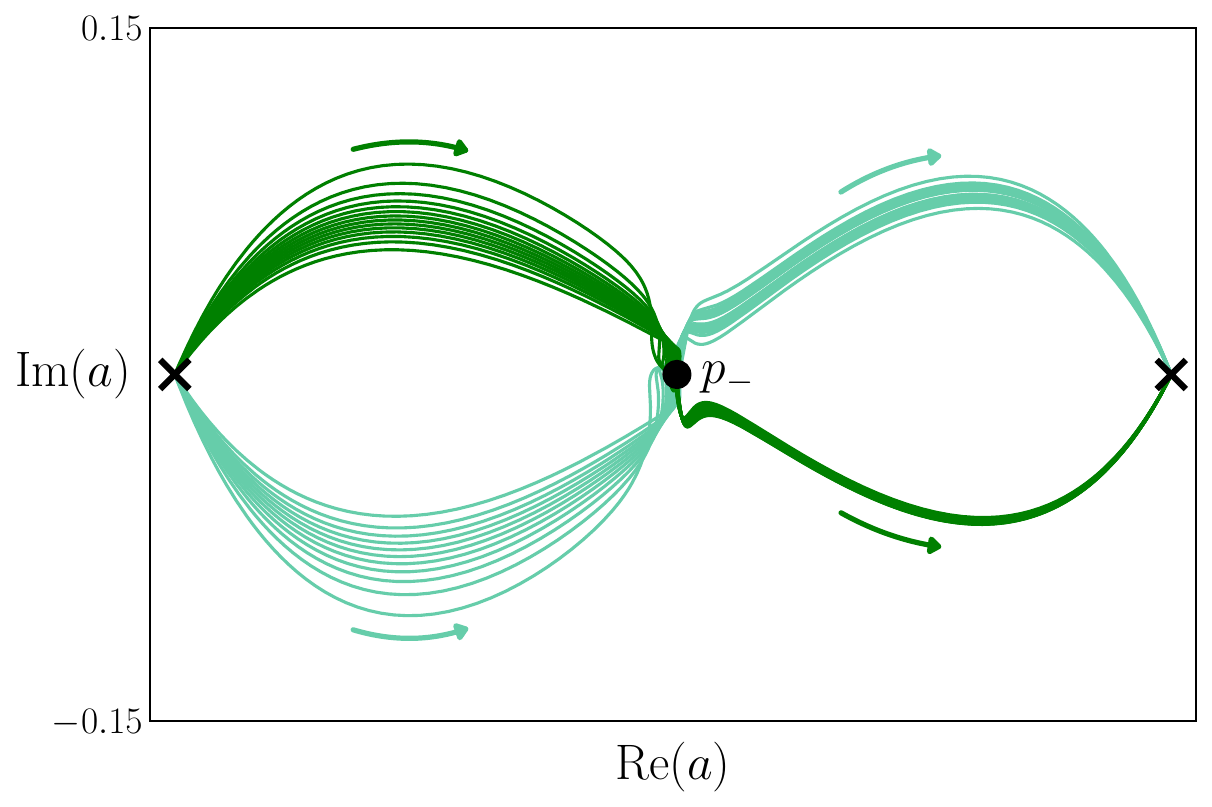}
    \includegraphics[width = 0.49\textwidth]{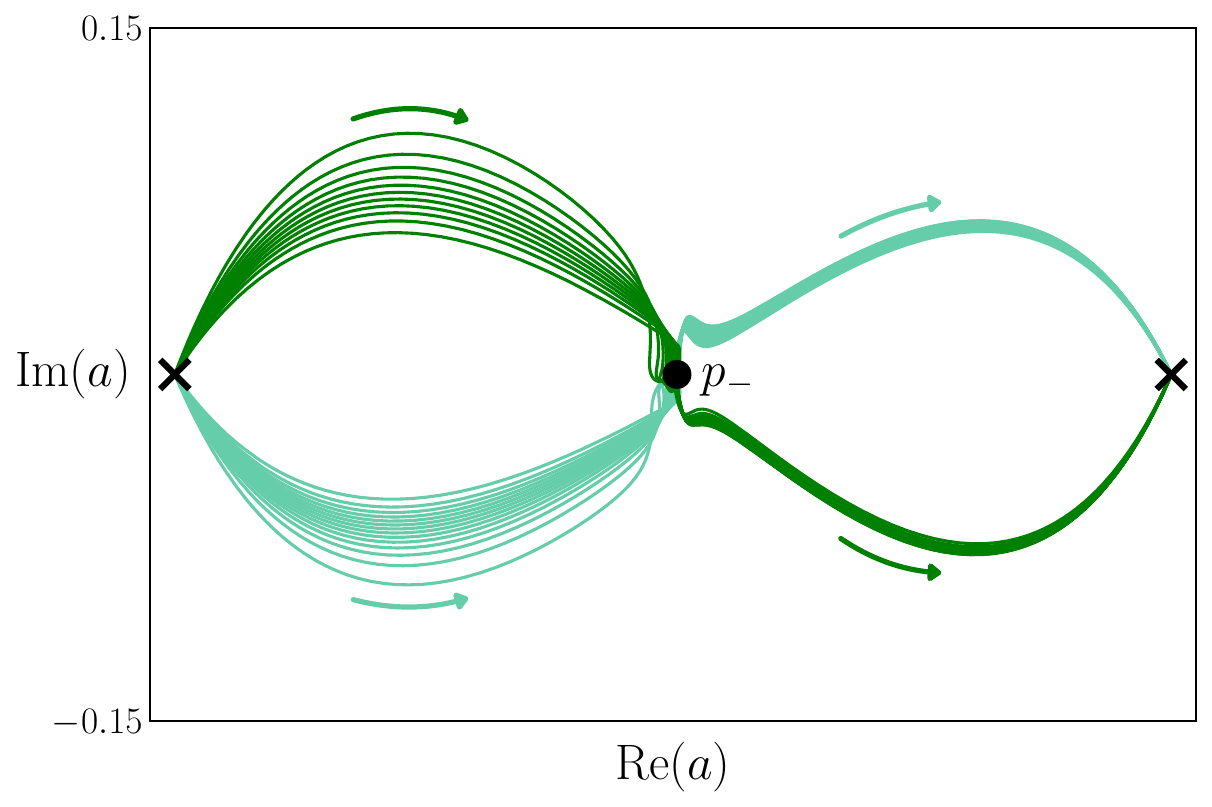}
    \includegraphics[width = 0.49\textwidth]{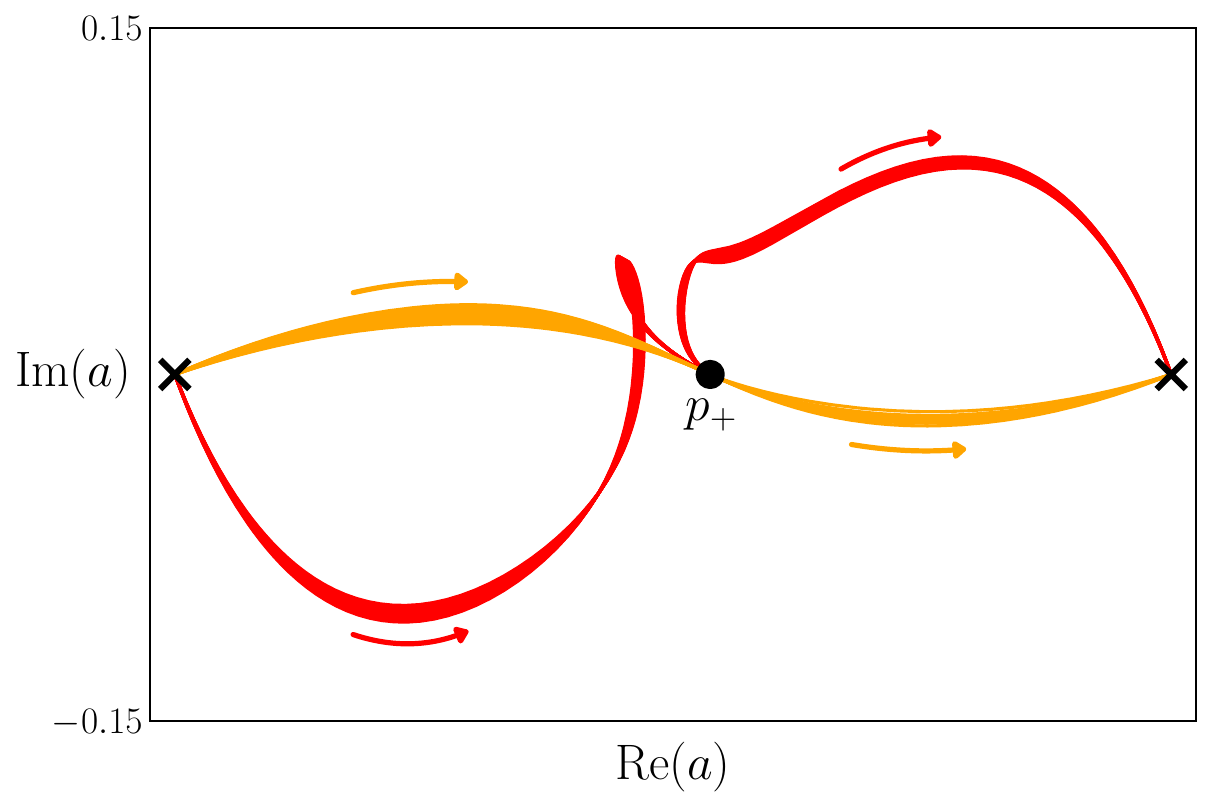}
    \includegraphics[width = 0.49\textwidth]{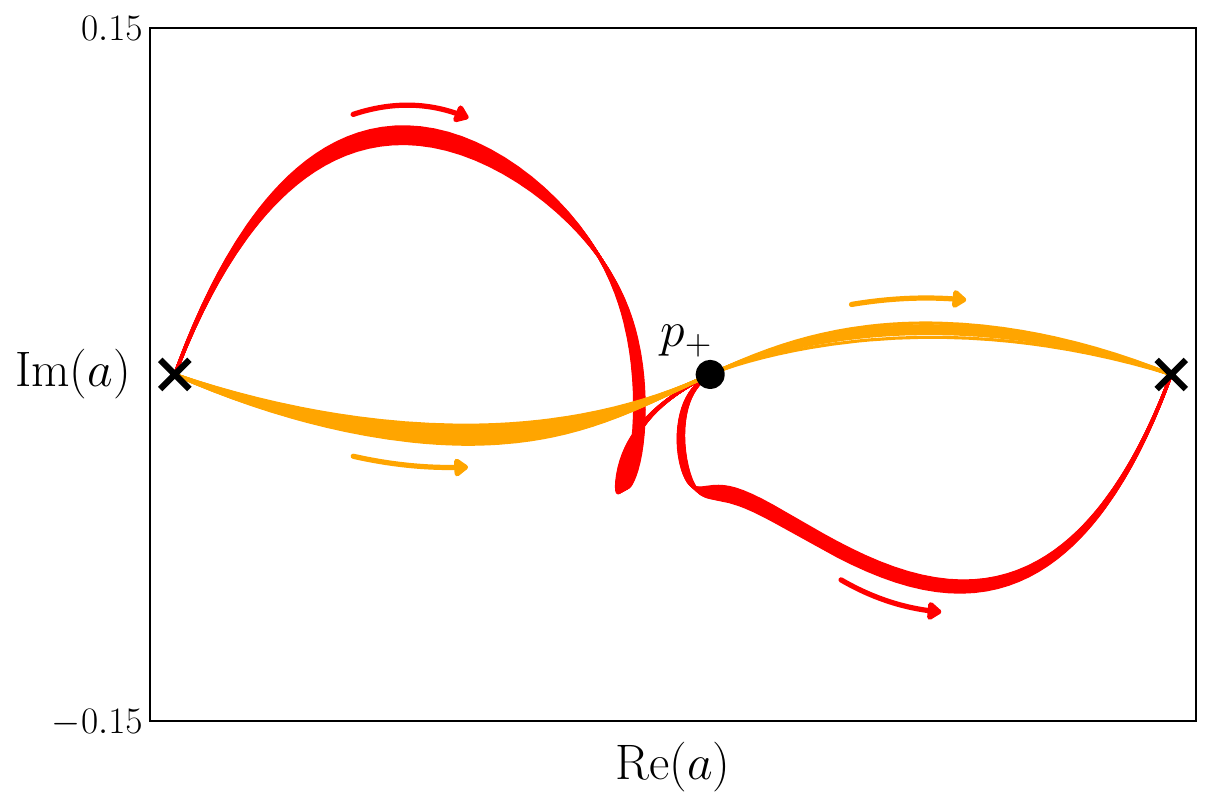}
    \caption{Projection on $a$--plane of downward flow lines from each saddle (shown separately) in the case $\beta = \frac{\pi}{2} < \beta_{\rm max}(\O)$ and $\O = 0.1$\,.
    The left panels correspond to $\Im(G) < 0$ and right panels correspond to $\Im(G) > 0$.
    Each saddle $p_{\sigma}$ is marked as a thick black dot.
    The flow lines asymptote towards $a = \pm 1$ (black cross) and are colour coded according to their asymptotic behaviour.}
    \label{fig: Projection on a plane}
\end{figure}

The thimble decomposition of $\mcc$ depends on the sign of $\Im(G)$, so we treat the two cases, $\Im(G) > 0$ and $\Im(G) < 0$, separately.
Since the thimbles are two-dimensional submanifolds of the full four-dimensional space, it is hard to visualize them directly.
Instead, we show the downward flow lines from each saddle by projecting them onto the complex $r_{+}$--plane in figure \ref{fig: Projection on rplus plane} and complex $a$--plane in figure \ref{fig: Projection on a plane}.

For saddle $p_{\sigma}$, we denote the thimble as $\mcj_{\sigma}^{-}$ in the case $\Im(G) < 0$ and as $\mcj_{\sigma}^{+}$ in the case $\Im(G) > 0$.
The thimble decomposition then reads as
\begin{empheq}[left = {\mcc = \empheqlbrace \,}]{align}
    \label{Decomposition 4}
    & \frac{1}{2}\mcj_{0}^{-} + \mcj_{-}^{-} + \mcj_{+}^{-}, \quad \Im(G) < 0, \\
    \label{Decomposition 5}
    & \frac{1}{2}\mcj_{0}^{+} - \mcj_{-}^{+} + \mcj_{+}^{+}, \quad \Im(G) > 0.
\end{empheq}
Even though the decompositions \eqref{Decomposition 4} and \eqref{Decomposition 5} differ, the resulting value of the integral is the same in the limit of vanishing $\Im(G)$.
To see this, we take a difference of the two integrals.
Since the mini-superspace integrand is continuous in the limit $\Im(G) \to 0$, we are left with its integral over a cycle which is homologous to zero.
This was expected since the original integral with $G$ real was absolutely convergent and well-defined.
In this way of thinking, the small black hole does contribute to the partition function for either sign of $\Im(G)$. \\

But, in the limit $\Im(G) \to 0$ (from either side) the small black hole contribution is purely imaginary and completely cancels off against the imaginary contribution coming from the thermal and large black hole saddles.
This is because the mini-superspace integral is itself real for a real value of $G$.
To make this cancellation manifest, we use the fact that with either sign of $\Im(G)$, the value of the integral is the same.
Further, the flow lines and hence the thimbles in the two cases are complex conjugates of each other.
So, if we add the two integrals, the two small black hole contributions as well as the imaginary contributions of the thermal and large black holes saddles cancel against each other, leaving behind the correct answer.
In this way of describing the integral for $\beta < \beta_{\rm max}(\O)$, we conclude that the the small black hole does not contribute. \\

Finally, we remark on another observation about the mini-superspace integral.
The derivatives of the exponent \eqref{Exponent 2} are polynomials in $r_{+}$ and $a$ divided by a power of $(1 - a^{2})$.
Each root of the polynomials with $a \neq 1$ constitutes a saddle for the integral.
It is clear that there are many saddles in the full complex plane of which we have only considered four, viz., thermal AdS and the Kerr-AdS black holes.
The ``extra'' saddles are also rotating black hole solutions but with a non-standard relation between temperature, rotation, and black hole parameters.
Fortunately, the decomposition of the integration region does not pick these additional saddles for $\beta > 0$ and $\abs{\O} < 1$.
Nevertheless, it would be interesting if any of these saddles, as well as the complex Kerr-AdS and spurious black hole saddles, are found directly from the gauge theory.


\section{Discussion}
\label{section: Discussion}
We posed an issue concerning the phase of holographic gauge theories at low temperatures and finite rotation.
The rotating AdS black holes exist as complex saddles at low temperatures with real part of on-shell action smaller than that of the thermal rotating gas.
Despite this, their contribution does not enter the partition function at any temperature and rotation.
We also addressed the question of whether the unstable small Kerr-AdS black hole contributes to the partition function at high temperatures.
It is worthwhile to ask whether saddles which do not contribute to the partition function can be seen at all from the gauge theory side.
An example of this is the map between small AdS black holes and small plasma balls in the confined phase \cite{Aharony:2005bm}. \\

The essential tool for our analysis was the mini-superspace approximation, which reduced the Euclidean path integral into a finite dimensional integral.
The residual integral was subjected to steepest descent analysis.
This approach has a limitation that it only gives qualitative information about the full path integral, such as which saddle points contribute in different ranges of the parameters.
For more detailed information, like the perturbative expansion around each saddle, we must go back to the full path integral.
Regardless, the simplicity of this approach makes it useful. \\

It would be nice if the simple methods presented in this paper can be used to discuss other novel situations.
A directly related example is understanding the CFT partition function at a fixed temperature and angular momentum from a bulk perspective.
The Hilbert space trace is certainly sensible in this case and the theory can exhibit different phases.
Another interesting example is the bulk path integral at a finite cutoff radius and modified boundary conditions \cite{DiTucci:2020weq,Banihashemi:2025qqi}.

\acknowledgments
I thank Raghu Mahajan and R. Loganayagam for various discussions and their comments on the draft.
I am grateful for numerous interactions with Gautam Mandal, Suvrat Raju, Ashoke Sen, and the rest of the ICTS string group.
I would also like to thank Ankur Barsode, Mayank Kumar Bijay, Ritwick Kumar Ghosh, Vinay Kumar, Aiswarya NS, Mithat \"Unsal for helping me with some resources.
I acknowledge support of the Department of Atomic Energy, Government of India, under project no. RTI4001.

\bibliography{main}

\providecommand{\href}[2]{#2}\begingroup\raggedright\begin{thebibliography}{10}

\bibitem{Bardeen:1973gs}
J.~M.~Bardeen, B.~Carter and S.~W.~Hawking, \emph{{The Four laws of black hole mechanics}}, \href{https://doi.org/10.1007/BF01645742}{\emph{Commun. Math. Phys.} {\bfseries 31} (1973) 161}.

\bibitem{Gibbons:1976ue}
G.~W.~Gibbons and S.~W.~Hawking, \emph{{Action Integrals and Partition Functions in Quantum Gravity}}, \href{https://doi.org/10.1103/PhysRevD.15.2752}{\emph{Phys. Rev. D} {\bfseries 15} (1977) 2752}.

\bibitem{Witten:2024upt}
E.~Witten, \emph{{Introduction to Black Hole Thermodynamics}},  \href{https://arxiv.org/abs/2412.16795}{{\ttfamily 2412.16795}}.

\bibitem{Hawking:1982dh}
S.~W.~Hawking and D.~N.~Page, \emph{{Thermodynamics of Black Holes in anti-De Sitter Space}}, \href{https://doi.org/10.1007/BF01208266}{\emph{Commun. Math. Phys.} {\bfseries 87} (1983) 577}.

\bibitem{Maldacena:1997re}
J.~M.~Maldacena, \emph{{The Large $N$ limit of superconformal field theories and supergravity}}, \href{https://doi.org/10.4310/ATMP.1998.v2.n2.a1}{\emph{Adv. Theor. Math. Phys.} {\bfseries 2} (1998) 231} [\href{https://arxiv.org/abs/hep-th/9711200}{{\ttfamily hep-th/9711200}}].

\bibitem{Gubser:1998bc}
S.~S.~Gubser, I.~R.~Klebanov and A.~M.~Polyakov, \emph{{Gauge theory correlators from noncritical string theory}}, \href{https://doi.org/10.1016/S0370-2693(98)00377-3}{\emph{Phys. Lett. B} {\bfseries 428} (1998) 105} [\href{https://arxiv.org/abs/hep-th/9802109}{{\ttfamily hep-th/9802109}}].

\bibitem{Witten:1998qj}
E.~Witten, \emph{{Anti de Sitter space and holography}}, \href{https://doi.org/10.4310/ATMP.1998.v2.n2.a2}{\emph{Adv. Theor. Math. Phys.} {\bfseries 2} (1998) 253} [\href{https://arxiv.org/abs/hep-th/9802150}{{\ttfamily hep-th/9802150}}].

\bibitem{Witten:1998zw}
E.~Witten, \emph{{Anti-de Sitter space, thermal phase transition, and confinement in gauge theories}}, \href{https://doi.org/10.4310/ATMP.1998.v2.n3.a3}{\emph{Adv. Theor. Math. Phys.} {\bfseries 2} (1998) 505} [\href{https://arxiv.org/abs/hep-th/9803131}{{\ttfamily hep-th/9803131}}].

\bibitem{Mahajan:2025bzo}
R.~Mahajan and K.~Singhi, \emph{{A Brief Note on Complex AdS-Schwarzschild Black Holes}},  \href{https://arxiv.org/abs/2509.08883}{{\ttfamily 2509.08883}}.

\bibitem{Jonas:2022uqb}
C.~Jonas, J.-L.~Lehners and J.~Quintin, \emph{{Uses of complex metrics in cosmology}}, \href{https://doi.org/10.1007/JHEP08(2022)284}{\emph{JHEP} {\bfseries 08} (2022) 284} [\href{https://arxiv.org/abs/2205.15332}{{\ttfamily 2205.15332}}].

\bibitem{Maldacena:2019cbz}
J.~Maldacena, G.~J.~Turiaci and Z.~Yang, \emph{{Two dimensional Nearly de Sitter gravity}}, \href{https://doi.org/10.1007/JHEP01(2021)139}{\emph{JHEP} {\bfseries 01} (2021) 139} [\href{https://arxiv.org/abs/1904.01911}{{\ttfamily 1904.01911}}].

\bibitem{Aharony:2003sx}
O.~Aharony, J.~Marsano, S.~Minwalla, K.~Papadodimas and M.~Van~Raamsdonk, \emph{{The Hagedorn - deconfinement phase transition in weakly coupled large N gauge theories}}, \href{https://doi.org/10.4310/ATMP.2004.v8.n4.a1}{\emph{Adv. Theor. Math. Phys.} {\bfseries 8} (2004) 603} [\href{https://arxiv.org/abs/hep-th/0310285}{{\ttfamily hep-th/0310285}}].

\bibitem{Hertog:2011ky}
T.~Hertog and J.~Hartle, \emph{{Holographic No-Boundary Measure}}, \href{https://doi.org/10.1007/JHEP05(2012)095}{\emph{JHEP} {\bfseries 05} (2012) 095} [\href{https://arxiv.org/abs/1111.6090}{{\ttfamily 1111.6090}}].

\bibitem{Feldbrugge:2017kzv}
J.~Feldbrugge, J.-L.~Lehners and N.~Turok, \emph{{Lorentzian Quantum Cosmology}}, \href{https://doi.org/10.1103/PhysRevD.95.103508}{\emph{Phys. Rev. D} {\bfseries 95} (2017) 103508} [\href{https://arxiv.org/abs/1703.02076}{{\ttfamily 1703.02076}}].

\bibitem{Narain:2021bff}
G.~Narain, \emph{{On Gauss-bonnet gravity and boundary conditions in Lorentzian path-integral quantization}}, \href{https://doi.org/10.1007/JHEP05(2021)273}{\emph{JHEP} {\bfseries 05} (2021) 273} [\href{https://arxiv.org/abs/2101.04644}{{\ttfamily 2101.04644}}].

\bibitem{Maldacena:2024uhs}
J.~Maldacena, \emph{{Comments on the no boundary wavefunction and slow roll inflation}},  \href{https://arxiv.org/abs/2403.10510}{{\ttfamily 2403.10510}}.

\bibitem{Ivo:2024ill}
V.~Ivo, Y.-Z.~Li and J.~Maldacena, \emph{{The no boundary density matrix}}, \href{https://doi.org/10.1007/JHEP02(2025)124}{\emph{JHEP} {\bfseries 02} (2025) 124} [\href{https://arxiv.org/abs/2409.14218}{{\ttfamily 2409.14218}}].

\bibitem{Turiaci:2025xwi}
G.~J.~Turiaci and C.-H.~Wu, \emph{{The wavefunction of a quantum S$^{1}$ {\texttimes} S$^{2}$ universe}}, \href{https://doi.org/10.1007/JHEP07(2025)158}{\emph{JHEP} {\bfseries 07} (2025) 158} [\href{https://arxiv.org/abs/2503.14639}{{\ttfamily 2503.14639}}].

\bibitem{Shi:2025amq}
X.~Shi and G.~J.~Turiaci, \emph{{The phase of the gravitational path integral}}, \href{https://doi.org/10.1007/JHEP07(2025)047}{\emph{JHEP} {\bfseries 07} (2025) 047} [\href{https://arxiv.org/abs/2504.00900}{{\ttfamily 2504.00900}}].

\bibitem{Ivo:2025yek}
V.~Ivo, J.~Maldacena and Z.~Sun, \emph{{Physical instabilities and the phase of the Euclidean path integral}},  \href{https://arxiv.org/abs/2504.00920}{{\ttfamily 2504.00920}}.

\bibitem{Saad:2018bqo}
P.~Saad, S.~H.~Shenker and D.~Stanford, \emph{{A semiclassical ramp in SYK and in gravity}},  \href{https://arxiv.org/abs/1806.06840}{{\ttfamily 1806.06840}}.

\bibitem{Saad:2019pqd}
P.~Saad, \emph{{Late Time Correlation Functions, Baby Universes, and ETH in JT Gravity}},  \href{https://arxiv.org/abs/1910.10311}{{\ttfamily 1910.10311}}.

\bibitem{Stanford:2020wkf}
D.~Stanford, \emph{{More quantum noise from wormholes}},  \href{https://arxiv.org/abs/2008.08570}{{\ttfamily 2008.08570}}.

\bibitem{Chen:2023hra}
Y.~Chen, V.~Ivo and J.~Maldacena, \emph{{Comments on the double cone wormhole}}, \href{https://doi.org/10.1007/JHEP04(2024)124}{\emph{JHEP} {\bfseries 04} (2024) 124} [\href{https://arxiv.org/abs/2310.11617}{{\ttfamily 2310.11617}}].

\bibitem{Skenderis:2008dh}
K.~Skenderis and B.~C.~van Rees, \emph{{Real-time gauge/gravity duality}}, \href{https://doi.org/10.1103/PhysRevLett.101.081601}{\emph{Phys. Rev. Lett.} {\bfseries 101} (2008) 081601} [\href{https://arxiv.org/abs/0805.0150}{{\ttfamily 0805.0150}}].

\bibitem{Glorioso:2018mmw}
P.~Glorioso, M.~Crossley and H.~Liu, \emph{{A prescription for holographic Schwinger-Keldysh contour in non-equilibrium systems}},  \href{https://arxiv.org/abs/1812.08785}{{\ttfamily 1812.08785}}.

\bibitem{Boruch:2023gfn}
J.~Boruch, L.~V.~Iliesiu, S.~Murthy and G.~J.~Turiaci, \emph{{New forms of attraction: attractor saddles for the black hole index}}, \href{https://doi.org/10.1007/JHEP04(2025)087}{\emph{JHEP} {\bfseries 04} (2025) 087} [\href{https://arxiv.org/abs/2310.07763}{{\ttfamily 2310.07763}}].

\bibitem{Hegde:2024bmb}
S.~Hegde, A.~Sen, P.~Shanmugapriya and A.~Virmani, \emph{{Supersymmetric index for half BPS black holes in N=2 supergravity with higher curvature corrections}}, \href{https://doi.org/10.1007/JHEP02(2025)131}{\emph{JHEP} {\bfseries 02} (2025) 131} [\href{https://arxiv.org/abs/2411.08260}{{\ttfamily 2411.08260}}].

\bibitem{Adhikari:2024zif}
S.~Adhikari, P.~Dharanipragada, K.~Goswami and A.~Virmani, \emph{{Attractor saddle for 5D black hole index}}, \href{https://doi.org/10.1007/JHEP03(2025)180}{\emph{JHEP} {\bfseries 03} (2025) 180} [\href{https://arxiv.org/abs/2411.12413}{{\ttfamily 2411.12413}}].

\bibitem{Boruch:2025biv}
J.~Boruch, L.~V.~Iliesiu, S.~Murthy and G.~J.~Turiaci, \emph{{Multicentered black hole saddles for supersymmetric indices}},  \href{https://arxiv.org/abs/2507.07166}{{\ttfamily 2507.07166}}.

\bibitem{Hawking:1998kw}
S.~W.~Hawking, C.~J.~Hunter and M.~Taylor, \emph{{Rotation and the AdS / CFT correspondence}}, \href{https://doi.org/10.1103/PhysRevD.59.064005}{\emph{Phys. Rev. D} {\bfseries 59} (1999) 064005} [\href{https://arxiv.org/abs/hep-th/9811056}{{\ttfamily hep-th/9811056}}].

\bibitem{Gibbons:2004ai}
G.~W.~Gibbons, M.~J.~Perry and C.~N.~Pope, \emph{{The First law of thermodynamics for Kerr-anti-de Sitter black holes}}, \href{https://doi.org/10.1088/0264-9381/22/9/002}{\emph{Class. Quant. Grav.} {\bfseries 22} (2005) 1503} [\href{https://arxiv.org/abs/hep-th/0408217}{{\ttfamily hep-th/0408217}}].

\bibitem{Skenderis:2002wp}
K.~Skenderis, \emph{{Lecture notes on holographic renormalization}}, \href{https://doi.org/10.1088/0264-9381/19/22/306}{\emph{Class. Quant. Grav.} {\bfseries 19} (2002) 5849} [\href{https://arxiv.org/abs/hep-th/0209067}{{\ttfamily hep-th/0209067}}].

\bibitem{Gibbons:2005jd}
G.~W.~Gibbons, M.~J.~Perry and C.~N.~Pope, \emph{{AdS/CFT Casimir energy for rotating black holes}}, \href{https://doi.org/10.1103/PhysRevLett.95.231601}{\emph{Phys. Rev. Lett.} {\bfseries 95} (2005) 231601} [\href{https://arxiv.org/abs/hep-th/0507034}{{\ttfamily hep-th/0507034}}].

\bibitem{Awad:2007me}
A.~M.~Awad, \emph{{First law, counterterms and Kerr-AdS(5) black hole}}, \href{https://doi.org/10.1142/S0218271809014522}{\emph{Int. J. Mod. Phys. D} {\bfseries 18} (2009) 405} [\href{https://arxiv.org/abs/0708.3458}{{\ttfamily 0708.3458}}].

\bibitem{Marolf:2022ybi}
D.~Marolf, \emph{{Gravitational thermodynamics without the conformal factor problem: partition functions and Euclidean saddles from Lorentzian path integrals}}, \href{https://doi.org/10.1007/JHEP07(2022)108}{\emph{JHEP} {\bfseries 07} (2022) 108} [\href{https://arxiv.org/abs/2203.07421}{{\ttfamily 2203.07421}}].

\bibitem{Witten:2010cx}
E.~Witten, \emph{{Analytic Continuation Of Chern-Simons Theory}}, {\emph{AMS/IP Stud. Adv. Math.} {\bfseries 50} (2011) 347} [\href{https://arxiv.org/abs/1001.2933}{{\ttfamily 1001.2933}}].

\bibitem{Dunne:2015eaa}
G.~V.~Dunne and M.~\"Unsal, \emph{{What is QFT? Resurgent trans-series, Lefschetz thimbles, and new exact saddles}}, \href{https://doi.org/10.22323/1.251.0010}{\emph{PoS} {\bfseries LATTICE2015} (2016) 010} [\href{https://arxiv.org/abs/1511.05977}{{\ttfamily 1511.05977}}].

\bibitem{Aharony:2005bm}
O.~Aharony, S.~Minwalla and T.~Wiseman, \emph{{Plasma-balls in large N gauge theories and localized black holes}}, \href{https://doi.org/10.1088/0264-9381/23/7/001}{\emph{Class. Quant. Grav.} {\bfseries 23} (2006) 2171} [\href{https://arxiv.org/abs/hep-th/0507219}{{\ttfamily hep-th/0507219}}].

\bibitem{DiTucci:2020weq}
A.~Di~Tucci, M.~P.~Heller and J.-L.~Lehners, \emph{{Lessons for quantum cosmology from anti{\textendash}de Sitter black holes}}, \href{https://doi.org/10.1103/PhysRevD.102.086011}{\emph{Phys. Rev. D} {\bfseries 102} (2020) 086011} [\href{https://arxiv.org/abs/2007.04872}{{\ttfamily 2007.04872}}].

\bibitem{Banihashemi:2025qqi}
B.~Banihashemi, E.~Shaghoulian and S.~Shashi, \emph{{Thermal effective actions from conformal boundary conditions in gravity}}, \href{https://doi.org/10.1088/1361-6382/adee72}{\emph{Class. Quant. Grav.} {\bfseries 42} (2025) 155004} [\href{https://arxiv.org/abs/2503.17471}{{\ttfamily 2503.17471}}].

\end{thebibliography}\endgroup

\end{document}